\documentclass[reprint,amsmath,amssymb,aps,]{revtex4-2}

\usepackage{tikz}
\usepackage{quantikz}
\usepackage{multirow}
\usepackage{graphicx}
\usepackage{dcolumn}
\usepackage{caption}
\usepackage{booktabs}
\usepackage{placeins}  
\usepackage{bm}
\usepackage{bbold}

\usepackage{algorithm}
\usepackage{mathtools}
\usepackage{algpseudocode}
\usepackage[T1]{fontenc}
\usepackage[caption=false]{subfig}
\usetikzlibrary{positioning, quotes}

\usepackage[backref=page]{hyperref}

\begin{document}
\preprint{APS/123-QED}

\title{Adaptive Graph Shrinking for Quantum Optimization of \\ Constrained Combinatorial Problems}

\author{Monit Sharma$^{1}$}
\author{Hoong Chuin Lau$^{1,2}$}
\email{Corresponding author email: hclau@smu.edu.sg}
\address{$^1$School of Computing and Information Systems,  Singapore Management University, Singapore}
\address{$^2$Institute of High Performance Computing, A*STAR, Singapore}

\begin{abstract}
A range of quantum algorithms, especially those leveraging variational parameterization and circuit-based optimization, are being studied as alternatives for solving classically intractable combinatorial optimization problems (COPs). However, their applicability is limited by hardware constraints, including shallow circuit depth, limited qubit counts, and noise. To mitigate these issues, we propose a hybrid classical--quantum framework based on \emph{graph shrinking} to reduce the number of variables and constraints in QUBO formulations of COPs, while preserving problem structure.

Our approach introduces three key ideas: (i) \emph{constraint-aware shrinking} that prevents merges that will likely violate problem-specific feasibility constraints, (ii) a \emph{verification-and-repair pipeline} to correct infeasible solutions post-optimization, and (iii) \emph{adaptive strategies} for recalculating correlations and controlling the graph shrinking process. We apply our approach to three standard benchmark problems---\emph{Multidimensional Knapsack (MDKP)}, \emph{Maximum Independent Set (MIS)}, and the \emph{Quadratic Assignment Problem (QAP)}.

Empirical results show that our approach improves solution feasibility, reduces repair complexity, and enhances quantum optimization quality on hardware-limited instances. These findings demonstrate a scalable pathway for applying near-term quantum algorithms to classically challenging constrained optimization problems.
\end{abstract}

\maketitle

\section{\label{sec:level_introduction}Introduction}
Combinatorial optimization problems (COPs) are fundamental in various fields, such as logistics \cite{juan2015review}, finance \cite{du1998handbook}, machine learning \cite{bengio2021machine}, etc. However, these problems are often NP-hard and computationally intractable for classical algorithms. Among COPs, the (weighted) Max-Cut problem stands out as a quintessential example of an NP-hard problem, where the goal is to partition the vertices of a weighted graph into two subsets such that the total weight of edges between the subsets is maximized. Despite its simple formulation, solving large-scale Max-Cut instances remains extremely challenging, often beyond the reach of current heuristic and exact solvers \cite{khot2007optimal}.

Recent advances in quantum computing have introduced promising approaches for solving COPs, such as the \textbf{Quantum Approximate Optimization Algorithm (QAOA)} \cite{farhi2014quantumapproximateoptimizationalgorithm} and the \textbf{Variational Quantum Eigensolver (VQE)} \cite{peruzzo2014variational}. These quantum algorithms are particularly well suited for solving quadratic unconstrained binary optimization (QUBO) problems, including Max-Cut. However, the scalability of these methods is severely limited by the number of qubits available on current quantum hardware, making it difficult to solve even medium-sized (not to mention large-scale) problems directly. 

For instance, IBM's Eagle processor features 127 qubits, and Rigetti's Aspen-M series offers around 80 qubits. However, due to factors such as gate fidelity, connectivity constraints, and error rates, the effective number of qubits usable for complex algorithms is often lower. Consequently, the largest weighted Max-Cut instances solvable without decomposition typically have fewer vertices than the nominal qubit count of the hardware \cite{sachdeva2024quantum}.

To overcome the limitations of current quantum hardware, we introduce a hybrid classical–quantum framework that employs graph shrinking as a principled method for constraint simplification and variable reduction. By iteratively merging structurally correlated vertices, our method reduces both the number of decision variables and the complexity of the constraints in QUBO formulations. This graph shrinking process systematically compresses the problem instance while preserving key structural properties of the original combinatorial model. As a result, it improves constraint propagation and reduces computational overhead during preprocessing, making the shrunk problem instances more amenable to quantum optimization. 

While the final step involves solving the reduced problem using a quantum simulator, the core contributions of this work lie in the classical components of the pipeline, namely adaptive graph shrinking, constraint-aware merging, and feasibility-preserving repair. Guided by correlations from Semi-Definite Programming (SDP) relaxations, our framework performs iterative vertex merging to compress large-scale combinatorial problems into smaller, tractable instances. Although this heuristic approach sacrifices exact optimality guarantees, it enables scalable execution on quantum backends by significantly reducing the problem size while preserving key structure property of the original problem.

\subsection{Overview and Contributions}

In contrast to prior work such as \cite{herzog2024improvingquantumclassicaldecomposition}, which applies graph shrinking primarily to aid circuit cutting via balanced separators, we present \emph{graph shrinking as a standalone preprocessing tool} to reduce both variable and constraint complexity before quantum optimization. Unlike the static policies in \cite{fischer2024quantumclassicalcorrelationsshrinking}, our approach incorporates \emph{adaptive correlation updates} and a \emph{spectral stopping criterion}, enabling dynamic and efficient control over the shrinking process.

To ensure feasibility and improve the robustness of quantum solutions for constrained combinatorial problems, we introduce a \textbf{dual constraint-handling strategy}:
\begin{itemize}
    \item A \textbf{Proactive Constraint-Aware Merging} mechanism that penalizes merges likely to violate structural constraints (e.g., adjacency in MIS, capacity in MDKP, or permutation in QAP);
    \item A \textbf{Reactive Verification-and-Repair} stage that applies lightweight, problem-specific heuristics to restore feasibility after quantum solution generation.
\end{itemize}

Together, these techniques form a unified hybrid classical–quantum framework that compresses complex instances into tractable subproblems, enabling effective application of quantum solvers such as QAOA and VQE. Our main contributions are:

\begin{enumerate}
    \item \textbf{A Hybrid Classical–Quantum Framework} for solving constrained combinatorial problems using adaptive graph shrinking and quantum solvers.
    
    \item \textbf{Adaptive Correlation Update Strategies}, including dynamic recalculation frequency and localized (partial) correlation updates, to improve scalability.
    
    \item \textbf{A Spectral-Based Shrinking Termination Criterion} that leverages eigenvalue trends of the graph Laplacian to halt shrinking while preserving structure.

    \item \textbf{Constraint-Aware Merging and Repair Mechanisms} that proactively prevent infeasibility and reactively fix violations, customized for our targetted problems of MIS, MDKP, and QAP.
\end{enumerate}

Note that while graph shrinking occurs most naturally on the Max-Cut problem, our framework is applicable to a wide range of COPs, since any QUBO formulation of a given problem can be transformed into an equivalent weighted Max-Cut problem by introducing one additional node~\cite{barahona1989experiments}.

In this paper, we apply our approach to solve the Multi-Dimensional Knapsack (MDKP) \cite{kellerer2004multidimensional}, the Maximum Independent Set (MIS)\cite{tarjan1977finding} and the Quadratic Assignment Problem (QAP) \cite{koopmans1957assignment}, and focus on solving classically proven hard benchmark instances of these problems (namely, \cite{drake2015benchmark,Sloane2000, Burkard1997QAPLIB,QAPLIB_Lehigh}), ensuring that our approach is tested on instances where classical methods struggle. 

The remainder of this paper is organized as follows: Section~\ref{sec:level1} provides a literature review on the other decomposition strategies used in optimization problems. Section~\ref{sec:qubo_formulations} presents the transformation of combinatorial optimization problems into QUBO formulations. We show how hard constraints are incorporated as quadratic penalty terms and discusses the role of penalty strength in ensuring that constraint violations are energetically unfavorable. 
Section~\ref{sec:qubo_to_maxcut} describes how the QUBO formulation is further transformed into a weighted Max-Cut problem. This section highlights how the dominance of constraint-related terms in the QUBO ensures that the SDP relaxation yields correlation patterns that accurately reflect the problem feasible structure, thereby guiding effective graph shrinking.

Sections~\ref{sec:level2} and \ref{sec:new-addition} introduce our key contributions, which is the graph shrinking methodology (Contributions 2 and 3) followed by our proposed approaches to address constraint feasbility (Contribution 4). Section~\ref{sec:level3} details the integration of graph shrinking with quantum optimization techniques. 

Section~\ref{sec:level7} describes the experimental setup, and the results and analysis. Finally, Section~\ref{sec:discussion} concludes with a discussion of the implications of our work and future research directions.

By combining classical graph shrinking techniques with quantum optimization, we aim to bridge the gap between classical and quantum computing, enabling scalable and efficient algorithms to be designed.

\section{\label{sec:level1}Literature Review}

In combinatorial optimization, decomposition techniques are fundamental to reducing problem complexity by partitioning large, intricate problems into smaller, more tractable subproblems. Such strategies are crucial in quantum optimization given limited qubit counts. For instance, Ponce et al.\cite{Ponce2023} decompose Max-Cut for QAOA into smaller subgraphs. Similarly, branch decomposition \cite{BranchDecomposition} clusters graph components to enable dynamic programming. Another emerging approach is warm-starting quantum algorithms with classical solutions, using, e.g., SDP relaxations to initialize QAOA states \cite{tate2022bridgingclassicalquantumsdp}.

In the context of the Traveling Salesman Problem (TSP), traditional approaches, such as branch-and-bound and cutting-plane methods, have leveraged problem-specific decompositions (e.g. \cite{padberg1991branch}). However, these methods are often problem-specific and may not generalize well to arbitrary COPs. An example is quantum walk-inspired state-space reduction \cite{StateSpaceReduction,harrigan2021quantum}, which constructs a superposition over all potential solution paths but prunes them by enforcing path spacing constraints. This focuses the quantum search on a smaller and relevant subspace. Furthermore, quantum-informed recursive optimization algorithms (QIRO) \cite{QIRO} harness quantum correlations to guide iterative classical updates, hinting that better quantum hardware would further enhance the performance of these methods.

The utility of geometric insights is further illustrated by the Planar Separator Theorem \cite{PlanarSeparator, lamm2016finding}, which guarantees that any planar graph can be partitioned by removing $\mathcal{O}(\sqrt{n})$ vertices, producing subgraphs with at most $2n/3$ vertices. This result has been instrumental in the design of quantum divide-and-conquer algorithms, such as those that achieve subexponential upper bounds in classical computation times for contracting tensor networks \cite{PhysRevLett.131.180601}.

Quantum annealing has also gained prominence in tackling combinatorial optimization challenges. Multilevel frameworks, such as the hybrid solvers introduced by Ushijima-Mwesigwa et al. \cite{ushijima2021multilevel}, integrate graph contraction with quantum local search across D-Wave and IBM quantum architectures to effectively address large-scale graph partitioning and community detection. Complementing these methods, a constraint programming approach for QUBO solving \cite{codognet2024constraint} employs logical inference to pre-solve or partition problems, thereby enhancing their compatibility with quantum annealing processes.

By acknowledging these classical methods, we clarify that our approach is not the first to use graph contractions. Rather, we propose a method that fits into this host of decomposition strategies.
Collectively, these approaches demonstrate how combining classical and quantum techniques can tackle the complexity of large-scale COPs. This interplay between classical reductions and quantum algorithms is key to making progress given current quantum hardware limitations.

\section{QUBO Formulations with Explicit Constraint Penalization}
\label{sec:qubo_formulations}

We detail the transformation of three benchmark constrained combinatorial optimization problems—Multi-Dimensional Knapsack Problem (MDKP),  Maximum Independent Set (MIS) and Quadratic Assignment Problem (QAP)—into unconstrained Quadratic Unconstrained Binary Optimization (QUBO) form. These reformulations encode hard constraints as additive penalty terms, enabling the use of quantum optimization methods such as QAOA, VQE, or quantum annealing. We include representative examples and analyze the role of penalty strength in preserving constraint feasibility during optimization.

\subsection{Multi-Dimensional Knapsack Problem (MDKP)}
\label{subsec:mdkp_qubo}

Given a profit vector $p \in \mathbb{R}^n$, and $m$ linear capacity constraints of the form $W x \leq C$, where $W \in \mathbb{R}^{m \times n}$, the MDKP is formulated as:
\begin{align}
    \max_{x \in \{0,1\}^n} \quad & \sum_{i=1}^n p_i x_i \\
    \text{s.t.} \quad & \sum_{i=1}^n w_{ji} x_i \leq C_j, \quad \forall j \in \{1, \dots, m\}.
\end{align}

This constrained 0-1 integer program is converted into QUBO form by relaxing each constraint through a quadratic penalty:
\begin{equation}
    \min_{x \in \{0,1\}^n} \; -\sum_{i=1}^n p_i x_i + \sum_{j=1}^m P_j \left( \sum_{i=1}^n w_{ji} x_i - C_j \right)^2.
    \label{eq:mdkp_qubo}
\end{equation}

To improve encoding fidelity, especially when $C_j$ is large or fractional, one can introduce binary slack variables $s_{jk} \in \{0,1\}$ with:
\begin{equation}
    \sum_{k=0}^{\kappa_j - 1} 2^k s_{jk} = C_j - \sum_{i=1}^n w_{ji} x_i, \quad \kappa_j = \lfloor \log_2 (C_j + 1) \rfloor,
    \label{eq:slack_equality}
\end{equation}
leading to a stricter constraint reformulation:
\begin{equation}
    P_j \cdot \left( \sum_{i=1}^n w_{ji} x_i + \sum_{k=0}^{\kappa_j - 1} 2^k s_{jk} - C_j \right)^2.
    \label{eq:slack_penalty}
\end{equation}

\paragraph{Worked Example.}
Consider a small MDKP with $n=3$ items and $m=1$ constraint:
\begin{align*}
    \text{Profits:} \quad & p = [5, 7, 4] \\
    \text{Weights:} \quad & w = [2, 3, 4] \\
    \text{Capacity:} \quad & C = 5
\end{align*}

Without slack variables, the QUBO formulation becomes:
\begin{align}
    \min_{x \in \{0,1\}^3} \; & -(5x_1 + 7x_2 + 4x_3) + P \cdot (2x_1 + 3x_2 + 4x_3 - 5)^2 \\
    = \; & x^\top Q x + c^\top x \quad \text{(expand to extract $Q$ and $c$)}
\end{align}

Here, $P$ is set based on $p_{\max} = 7$, e.g., $P = 70$ ensures constraint violations are heavily penalized.

\paragraph{Penalty Strength.}
Following \cite{lucas2014ising}, penalty weights $P_j$ should be selected to exceed the maximum gain from violating a constraint, typically:
\begin{equation}
    P_j \geq \lambda \cdot \max_i p_i, \quad \lambda \in [10,100].
\end{equation}
This ensures feasible solutions are energetically favored.

\subsection{Maximum Independent Set (MIS)}
\label{subsec:mis_qubo}

Given a graph $G = (V,E)$, the MIS problem seeks the largest subset $S \subseteq V$ such that no two vertices in $S$ share an edge. Letting $x_i \in \{0,1\}$ encode membership in $S$, the original problem is:
\begin{align}
    \max_{x \in \{0,1\}^n} \quad & \sum_{i \in V} x_i \\
    \text{s.t.} \quad & x_i + x_j \le 1, \quad \forall (i,j) \in E.
\end{align}

To convert into QUBO form, we penalize violations of the independence constraint via pairwise quadratic terms:
\begin{equation}
    \min_{x \in \{0,1\}^n} \; -\sum_{i \in V} x_i + P \sum_{(i,j) \in E} x_i x_j.
    \label{eq:mis_qubo}
\end{equation}

\paragraph{Worked Example.}
Consider a triangle graph with vertices $V = \{1,2,3\}$ and edges $E = \{(1,2), (2,3), (1,3)\}$. The QUBO becomes:
\begin{equation}
    \min_{x \in \{0,1\}^3} \; -(x_1 + x_2 + x_3) + P (x_1x_2 + x_2x_3 + x_1x_3)
\end{equation}

The optimal independent set is any single vertex, as selecting two incurs a penalty of $P$. For $P=2$, the minimal solution is $\{x_1 = 1, x_2 = x_3 = 0\}$, with objective $-1$.

\paragraph{Penalty Strength.}
To discourage selecting adjacent vertices:
\begin{equation}
    P > 1 \quad \text{(for unweighted MIS)},
\end{equation}
ensuring that the penalty outweighs the gain of selecting two adjacent nodes. In practice, we set:
\begin{equation}
    P = \lambda, \quad \lambda \in [3,10],
\end{equation}
scaling with graph density or in proportion to the number of edges.

\subsection{Quadratic Assignment Problem (QAP)}
\label{subsec:qap_qubo}

The Quadratic Assignment Problem (QAP) models the task of assigning $n$ facilities to $n$ locations such that the total cost, determined by the flow between facilities and the distance between locations, is minimized. Let $F \in \mathbb{R}^{n \times n}$ denote the flow matrix between facilities and $D \in \mathbb{R}^{n \times n}$ the distance matrix between locations. Define binary variables $x_{ij} \in \{0,1\}$, where $x_{ij} = 1$ if facility $i$ is assigned to location $j$.

The classical formulation is:
\begin{align}
    \min_{x_{ij} \in \{0,1\}} \quad & \sum_{i=1}^n \sum_{j=1}^n \sum_{k=1}^n \sum_{l=1}^n F_{ik} D_{jl} x_{ij} x_{kl} \\
    \text{s.t.} \quad & \sum_{j=1}^n x_{ij} = 1, \quad \forall i \in \{1, \dots, n\} \\
    & \sum_{i=1}^n x_{ij} = 1, \quad \forall j \in \{1, \dots, n\}.
\end{align}

Each facility must be assigned to exactly one location, and vice versa. To convert this into a QUBO form, we flatten the $n \times n$ matrix $x$ into a binary vector $x \in \{0,1\}^{n^2}$, and rewrite the constraints as quadratic penalties:
\begin{equation}
    \min_{x \in \{0,1\}^{n^2}} \; x^\top Q x + \sum_{i=1}^n P_i \left( \sum_{j=1}^n x_{ij} - 1 \right)^2 + \sum_{j=1}^n P_j \left( \sum_{i=1}^n x_{ij} - 1 \right)^2.
    \label{eq:qap_qubo}
\end{equation}

The quadratic term $x^\top Q x$ encodes the objective function, with $Q$ constructed such that each entry $Q_{(i,j),(k,l)} = F_{ik} D_{jl}$. The penalty terms ensure that each facility and each location are uniquely matched.

\paragraph{Worked Example.} For a simple QAP with $n=2$:
\begin{align*}
    \text{Flow:} & \quad F = \begin{bmatrix} 0 & 5 \\ 5 & 0 \end{bmatrix} \\
    \text{Distance:} & \quad D = \begin{bmatrix} 0 & 2 \\ 2 & 0 \end{bmatrix}
\end{align*}
Let $x = [x_{11}, x_{12}, x_{21}, x_{22}]^\top$. The objective becomes:
\begin{equation}
    x^\top Q x = 20(x_{11}x_{22} + x_{12}x_{21})
\end{equation}
Subject to the constraints:
\begin{align*}
    x_{11} + x_{12} &= 1 \\
    x_{21} + x_{22} &= 1 \\
    x_{11} + x_{21} &= 1 \\
    x_{12} + x_{22} &= 1
\end{align*}
These are penalized quadratically and added to the QUBO objective as in Equation~\eqref{eq:qap_qubo}.

\paragraph{Penalty Strength.} Following \cite{lucas2014ising}, the penalty weight $P$ must exceed the maximum contribution of any feasible objective term:
\begin{equation}
    P \geq \lambda \cdot \max_{i,k,j,l} |F_{ik} D_{jl}|, \quad \lambda \in [10, 100].
\end{equation}
This guarantees that constraint-violating assignments are energetically disfavored during optimization.

These QUBO encodings transform constrained 0-1 integer programs into unconstrained quadratic forms suitable for quantum optimization. Following established conventions \cite{lucas2014ising, lechner2020quantum}, penalty coefficients must be chosen to be both problem-specific and empirically validated. Too weak a penalty leads to infeasible solutions; too strong a penalty may flatten the optimization landscape, increasing convergence difficulty. Our formulation ensures a robust balance, and supports systematic transformation of domain-constrained COPs into a unified QUBO framework for downstream processing via Max-Cut relaxation and graph contraction.

\section{Transformation from QUBO to Weighted Max-Cut}
\label{sec:qubo_to_maxcut}

In order to apply correlation-guided graph shrinking to arbitrary constrained optimization problems, we must first transform the associated QUBO formulation into an equivalent unconstrained weighted Max-Cut problem. This section clarifies the transformation mechanics, particularly how constraint-derived penalty terms in the QUBO affect the structure of the Max-Cut graph and its associated correlations obtained via semi-definite programming (SDP) relaxation.

\subsection{Barahona Reduction: QUBO to Max-Cut}
\label{subsec:barahona}

Let the original QUBO be represented as:
\begin{equation}
    H(x) = x^\top Q x + c^\top x, \quad x \in \{0,1\}^n,
\end{equation}
where \( Q \in \mathbb{R}^{n \times n} \) is a symmetric matrix and \( c \in \mathbb{R}^n \) is a linear coefficient vector. We follow the well-established reduction by Barahona et al.~\cite{barahona1989experiments} which maps this QUBO to a weighted Max-Cut instance on \( n+1 \) nodes.

The mapping proceeds by introducing an auxiliary node \( 0 \) (called the reference node), and constructing a graph \( G' = (V', E') \) with node set \( V' = \{0, 1, 2, \dots, n\} \). Each QUBO variable \( x_i \) is associated with a binary spin variable \( z_i \in \{-1, +1\} \) via the transformation:
\[
x_i = \frac{1 - z_i}{2}.
\]

Substituting this into the QUBO, the resulting function over spin variables becomes a quadratic form that can be interpreted as a weighted Max-Cut Hamiltonian:
\begin{equation}
    H(z) = \sum_{i<j} w_{ij} \cdot \frac{1 - z_i z_j}{2} + \sum_{i} w_{0i} \cdot \frac{1 - z_0 z_i}{2},
    \label{eq:maxcut_form}
\end{equation}
where the edge weights \( w_{ij} \) and \( w_{0i} \) are defined by:
\begin{align}
    w_{ij} &= 4 Q_{ij} \quad \text{for } i < j, \\
    w_{0i} &= 2 Q_{ii} + c_i.
\end{align}

Hence, every entry of the QUBO matrix \( Q \) and linear term \( c \) directly contributes to edge weights in the Max-Cut graph, including those originating from constraint-penalty terms.

\subsection{Propagation of Constraint Penalties to Max-Cut Graph}
\label{subsec:constraint_influence}

In a constrained QUBO, penalty terms enforce feasibility. For example, the MDKP constraint:
\[
\left( \sum_{i=1}^n w_{ji} x_i - C_j \right)^2 = \sum_{i,k} w_{ji} w_{jk} x_i x_k - 2C_j \sum_i w_{ji} x_i + C_j^2
\]
adds structured quadratic and linear components to the QUBO matrix. These terms are typically scaled by large penalty coefficients \( P_j \), resulting in substantial entries in both \( Q \) and \( c \). Upon reduction, these large entries become dominant edge weights \( w_{ij} \) and \( w_{0i} \) in the Max-Cut graph.

We therefore categorize edges in the Max-Cut graph into two classes:
\begin{enumerate}
    \item \textbf{Objective edges:} Induced by the original linear and quadratic terms of the optimization objective (e.g., profits in MDKP, set size in MIS).
    \item \textbf{Constraint edges:} Induced by the penalty terms designed to enforce feasibility.
\end{enumerate}

To ensure that constraint satisfaction remains influential during graph shrinking, the following condition should hold:
\begin{equation}
    \min_{\text{constraint edges}} |w_{ij}| \gg \max_{\text{objective edges}} |w_{ij}|.
\end{equation}

This guarantees that high-penalty constraint-derived edges yield strong correlations during SDP relaxation and are preferentially preserved during node contraction. The SDP solver, when minimizing the relaxed Max-Cut Hamiltonian, will favor spin alignments (i.e., correlation patterns) that satisfy constraints, as violating them incurs a larger energetic cost.

\subsection{Effect on SDP-Derived Correlations}
\label{subsec:sdp_effect}

Let \( X \in \mathbb{R}^{(n+1) \times (n+1)} \) be the solution to the Max-Cut SDP relaxation~\cite{goemans1995improved}. The off-diagonal entries \( X_{ij} \) represent approximate spin correlations: 
\[
C_{ij} := X_{ij} \approx \mathbb{E}[z_i z_j] \in [-1, 1].
\]

Constraint-derived edges, having larger weights \( w_{ij} \), exert stronger influence on the SDP objective and tend to yield correlations \( C_{ij} \) close to \( \pm 1 \). Consequently, these edges dominate the merging decisions in our graph shrinking strategy:
\begin{equation}
(i,j) = \arg \max_{(i,j)} |C_{ij}|, \quad \text{merge if } |C_{ij}| \geq \tau.
\end{equation}

Therefore, penalty-enforced structure from the original QUBO—such as disallowing two connected vertices in MIS, or enforcing knapsack capacities in MDKP—remains visible and dominant in the SDP correlation matrix.

\subsection{Implication for Shrinking Heuristic}
\label{subsec:implication_shrinking}

By preserving constraint-derived edge dominance through appropriately scaled penalties, our framework ensures that the SDP-based contraction procedure merges variables in a way that implicitly respects original problem constraints. However, this is heuristic in nature and cannot formally guarantee feasibility. Hence, as described in Section~\ref{sec:new-addition}, we incorporate post-reconstruction feasibility checks and repair procedures to enforce constraint validity.

This transformation pipeline— of QUBO $\rightarrow$ Max-Cut $\rightarrow$ Shrinking via SDP correlations, thus retains the semantic structure of constraint satisfaction within the Max-Cut objective, making it suitable for constraint-aware preprocessing for quantum optimization.

\section{\label{sec:level2}Graph Shrinking Methodology}

Graph shrinking is a technique for reducing the size of a graph while preserving its essential structural properties, thereby improving the tractability of large-scale optimization problems. By iteratively merging vertices based on criteria such as correlation or edge weights, this method constructs a smaller graph that retains the key characteristics of the original. This approach is particularly effective for the MAXCUT problem, as it enables efficient computation while ensuring accurate solution reconstruction.

The methodology consists of several key steps: initialization, correlation analysis, thresholding, vertex merging, and solution reconstruction. Each step is designed to enhance scalability and computational efficiency, making the technique well-suited for large and complex graph-based optimization tasks. For a detailed explanation of the calculations and procedural steps, refer to \cite{fischer2024quantumclassicalcorrelationsshrinking}.

Since our approach employs semidefinite programming (SDP) correlation methods, we provide a brief explanation of their role, along with the adaptive recalculation strategy and the criteria for determining the number of nodes to which the graph is reduced.

\subsection{Computing Correlations Using the SDP Method}

The shrinking procedure begins by calculating correlations between the variables (nodes) of the graph. These correlations are essential for guiding the graph's simplification process. When employing the Semi-Definite Programming (SDP) method, the approach leverages a relaxation of the MAXCUT problem, enabling efficient computation of correlations.

\paragraph{SDP Relaxation Formulation:}
The MAXCUT problem can be mathematically expressed as:
\begin{equation}
C(x) = \frac{1}{2} \sum_{ij \in E} w_{ij} (1 - x_i x_j),
\end{equation}
where \( x \in \{-1, 1\}^n \) are binary variables indicating the partitions of the graph, \( w_{ij} \) are the edge weights, and \( E \) represents the set of edges.

To relax the problem, the binary variables \( x_i \) are replaced with unit vectors \( \mathbf{v}_i \in \mathbb{R}^n \), ensuring \( \|\mathbf{v}_i\| = 1 \). The relaxed problem is then formulated as:
\begin{equation}
\max_{\{\mathbf{v}_i\}} \frac{1}{2} \sum_{ij \in E} w_{ij} (1 - \mathbf{v}_i \cdot \mathbf{v}_j).
\end{equation}

\paragraph{Matrix Representation:}
This relaxation can also be represented in matrix form:
\begin{equation}
\max_{X} \left\{ \frac{1}{4} \langle L, X \rangle : \text{diag}(X) = \mathbf{e}, \, X \succeq 0 \right\},
\end{equation}
where:
\begin{itemize}
    \item \( X \) is the Gram matrix of the vectors \( \{\mathbf{v}_i\} \), such that \( X_{ij} = \mathbf{v}_i \cdot \mathbf{v}_j \),
    \item \( L \) is the Laplacian matrix of the graph, defined as \( L = \text{diag}(A \mathbf{e}) - A \), where \( A \) is the adjacency matrix,
    \item \( \langle L, X \rangle \) is the Frobenius inner product, expressed as \( \text{tr}(L^\top X) \),
    \item \( \text{diag}(X) = \mathbf{e} \) enforces that \( \|\mathbf{v}_i\| = 1 \),
    \item \( X \succeq 0 \) ensures that \( X \) is positive semi-definite.
\end{itemize}

\paragraph{Extracting Correlations:}
Once the SDP relaxation is solved, correlations are directly extracted from the Gram matrix \( X \). The correlation between any two nodes \( i \) and \( j \) is given by:
\begin{equation}
b_{ij}^{\text{SDP}} = \mathbf{v}_i \cdot \mathbf{v}_j = X_{ij}.
\end{equation}

The properties of \( b_{ij}^{\text{SDP}} \) are as follows:
\begin{itemize}
    \item \( b_{ij}^{\text{SDP}} \in [-1, 1] \),
    \item \( b_{ij}^{\text{SDP}} \approx 1 \) indicates strong alignment, suggesting that nodes \( i \) and \( j \) belong to the same partition,
    \item \( b_{ij}^{\text{SDP}} \approx -1 \) indicates anti-alignment, suggesting that nodes \( i \) and \( j \) belong to opposite partitions.
\end{itemize}

\paragraph{Key Insights:}
\begin{itemize}
    \item SDP correlations provide fractional values that effectively capture the alignment between nodes.
    \item These correlations prioritize the strongest edges, guiding the shrinking procedure.
    \item The SDP relaxation enables efficient computation and provides high-quality approximations for the MAXCUT problem.
\end{itemize}

\begin{figure}

\centering

\begin{tikzpicture}[remember picture, node distance=1cm]

\node[anchor=center] (A) {
    \subfloat[Initial graph with 4 nodes]{
        \begin{tikzpicture}[scale=0.8]
        
        \node[circle, draw,fill=blue!30] (1) at (0,0) {1};
        \node[circle, draw,fill=red!30] (2) at (2,0) {2};
        \node[circle, draw,fill=red!30] (4) at (2,2) {4};
        \node[circle, draw,fill=red!30] (3) at (0,2) {3};
        
        \draw (1) -- node[above] {5} (2);
        \draw (1) -- node[left] {2} (3);
        \draw (2) -- node[right] {3} (4);
        \draw (3) -- node[above] {4} (4);
        \coordinate (a_center) at (current bounding box.center);
        \end{tikzpicture}
    }
};

\node[anchor=center, right=3cm of A] (B) {
    \subfloat[Graph after reduction to 3 nodes]{
        \begin{tikzpicture}[scale=0.8]
        
        \node[circle, draw,fill=red!30] (2) at (2,0) {2};
        \node[circle, draw,fill=blue!30] (3) at (0,2) {3};
        \node[circle, draw,fill=red!30] (4) at (2,2) {4};
        
        \draw (2) -- node[above] {7} (3);
        \draw (2) -- node[right] {3} (4);
        \draw (3) -- node[above] {4} (4);
        \coordinate (b_center) at (current bounding box.center);
        \end{tikzpicture}
    }
};

\node[anchor=center, below=3cm of B] (C) {
    \subfloat[Final graph with 2 nodes]{
        \begin{tikzpicture}[scale=0.8]
        
        \node[circle, draw,fill=red!30] (2) at (0,0) {2};
        \node[circle, draw,fill=red!30] (4) at (2,0) {4};
        
        \draw (2) -- (4);
        \coordinate (c_center) at (current bounding box.center);
        \end{tikzpicture}
    }
};

\draw[->, thick] (A.east) -- (B.west);
\draw[->, thick] (B.south) -- (C.north);

\end{tikzpicture}

\caption{Sequence of graph reductions. 
Starting with the initial graph $V = \{1, 2, 3, 4\}$, we iteratively reduce its size based on correlation values. The correlation matrix $C = \begin{bmatrix}
1 & 0.9 & 0.3 & -0.2 \\
0.9 & 1 & 0.5 & -0.8 \\
0.3 & 0.5 & 1 & 0.7 \\
-0.2 & -0.8 & 0.7 & 1
\end{bmatrix}.$ guides the reduction process, where the strongest correlation $b_{12} = 0.9$ leads to merging nodes $1$ and $2$, reducing the graph to $V = \{2, 3, 4\}$ and finally to $V = \{2, 4\}$.}

\label{fig:graph_reductions}

\end{figure}
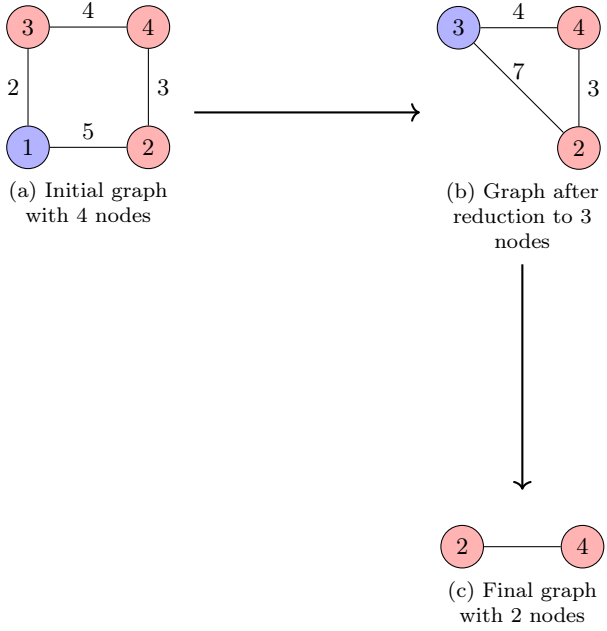

\subsection{Shrinking Step}

The shrinking step simplifies the graph by iteratively reducing its size while preserving the structural consistency of the original problem. The detailed steps involved in this procedure are as follows:

\paragraph{Selecting the Strongest Correlation:} 

The correlation matrix \( C \) guides the graph shrinking process. The algorithm identifies the edge \( (i, j) \) with the largest absolute correlation \( |b_{ij}| \). In the case of multiple edges having the same correlation value, ties are broken randomly.

The sign of the correlation \( b_{ij} \) determines the relationship between the nodes:
\begin{equation}
\sigma_{ij} = \text{sign}(b_{ij}).
\end{equation}
\begin{itemize}
    \item \( \sigma_{ij} = +1 \): Nodes \( i \) and \( j \) are assigned to the same partition.
    \item \( \sigma_{ij} = -1 \): Nodes \( i \) and \( j \) are assigned to opposite partitions.
\end{itemize}

\paragraph{Updating the Graph:}

After determining the relationship between nodes \( i \) and \( j \), the graph is updated to reflect their merging:

\begin{itemize}
    \item Node \( i \) is removed, and its edges are reassigned to node \( j \).
    \item For every neighbor \( k \) of \( i \) or \( j \), the edge weights are updated as:
    \begin{equation}
    w_{jk} =
    \begin{cases} 
    w_{jk} + \sigma_{ij} w_{ik}, & \text{if } jk \in E, \\
    \sigma_{ij} w_{ik}, & \text{if } jk \notin E.
    \end{cases}
    \end{equation}
\end{itemize}

\paragraph{Special Cases:}

If either \( i \) or \( j \) has already been merged with other nodes in previous steps, care must be taken to compute the effective correlation:
\begin{itemize}
    \item Use the previously computed correlations to determine the effective \( b_{ij} \) for the merged nodes.
    \item Skip correlations involving nodes that have already been fully merged, and proceed to the next strongest correlation.
\end{itemize}

\paragraph{Recalculation of Correlations:}

To ensure the relevance of the correlations to the modified graph structure, the correlations are recomputed every \( r \) steps. This dynamic recalculation adapts the process to the evolving graph structure. The value of \( r \) can be tuned based on the problem size and complexity. A smaller \( r \) ensures higher accuracy but increases computational cost.

Building on this, more sophisticated strategies for recalculating correlations can further optimize the graph shrinking process:

\subparagraph{1. Adaptive Recalculation Frequency:}  
Instead of recalculating after a fixed number of steps (\( r \)), the frequency is dynamically adjusted based on the state of the graph:
\begin{itemize}
    \item \textit{Change Detection:} Correlations are recalculated only when there is a significant structural change in the graph. For instance, if the number of edges decreases substantially:
    \[
    \Delta |E| = \frac{|E_{\text{current}}| - |E_{\text{previous}}|}{|E_{\text{previous}}|} > \delta,
    \]
    where \( \delta \) is a user-defined threshold. This ensures computational effort is focused on scenarios where the graph's topology changes significantly, potentially invalidating existing correlations.

    \item \textit{Threshold-Based Trigger:} Monitor the largest absolute correlation value \( |C_{ij}| \) in the matrix. If the maximum correlation drops below a predefined threshold (\( \tau \)):
    \[
    \max_{(i,j)} |C_{ij}| < \tau,
    \]
    recalculation is triggered. This strategy dynamically adapts to the graph's state, recalculating only when correlations weaken significantly.
\end{itemize}

\subparagraph{2. Partial Recalculation:}  
Full recalculation of the correlation matrix can be avoided by focusing only on affected regions of the graph:
\begin{itemize}
    \item \textit{Local Update:} When two nodes \( i \) and \( j \) are merged, update correlations involving their neighbors instead of recalculating the entire matrix. For neighbors \( k \in \mathcal{N}(i) \cup \mathcal{N}(j) \):
    \[
    C'_{jk} = \frac{w_{jk} + \sigma_{ij} w_{ik}}{\sqrt{d_j \cdot d_k}},
    \]
    where \( \sigma_{ij} = \text{sign}(C_{ij}) \), and \( d_j \), \( d_k \) are the degrees of nodes \( j \) and \( k \), respectively. This local adjustment minimizes computation while maintaining accuracy.

    \item \textit{Sparse Matrix Techniques:} Represent the correlation matrix as a sparse matrix, storing only nonzero entries. After merging nodes \( i \) and \( j \), update only the entries involving their neighbors. This method leverages sparsity to focus computations on meaningful entries, reducing overhead.
\end{itemize}

These advanced recalculation strategies strike a balance between computational efficiency and accuracy, enabling the graph shrinking algorithm to scale effectively for larger and more complex problems.\\

\paragraph{Stopping Condition:}

The shrinking process continues until the graph is reduced to the desired number of nodes, either specified by the user or determined intrinsically based on the graph's spectral properties. By leveraging spectral analysis, we can identify a natural threshold for reduction, ensuring that the essential structural and optimization properties of the original graph are preserved. This adaptive approach prevents excessive loss of critical information while maintaining computational efficiency, making it particularly suitable for large-scale optimization problems. A detailed explanation of this method, including the spectral criteria used for node selection and the steps involved in the reduction process, is provided in the following:

\subparagraph{1. Spectral Method for Determining Target Number of Nodes:}  

The stopping criterion in our graph shrinking framework can be governed either by a user-specified target node count \(k\) or determined dynamically through spectral analysis. Both approaches offer distinct trade-offs between control, structure preservation, and hardware compatibility.

\begin{itemize}

\item \textit{User-Specified Target Size.}  
Users may directly specify the number of nodes \(k\) to reduce to, typically based on the number of qubits available on the quantum hardware or based on empirical thresholds derived from solver performance. For instance, if the downstream quantum solver supports up to 64 qubits, one may choose \(k = 60\) to allow room for ancilla or auxiliary qubits. This approach ensures compatibility with target hardware but does not inherently preserve structural fidelity.

\item \textit{Spectral Energy Retention Criterion.}  
To preserve the structural essence of the graph, we employ a spectral method based on the eigenvalues of the Laplacian matrix. Let \(\lambda_1 \leq \lambda_2 \leq \dots \leq \lambda_n\) denote the eigenvalues of the (unnormalized) Laplacian. The cumulative retained energy for the top \(k\) eigenmodes is computed as:
\[
\text{Energy}_k = \frac{\sum_{i=1}^k \lambda_i}{\sum_{i=1}^n \lambda_i}.
\]
We choose the smallest \(k\) such that \(\text{Energy}_k \geq \alpha\), where \(\alpha \in [0,1]\) is a user-defined threshold (e.g., \(\alpha = 0.90\) for 90\% energy retention).

This method is inspired by the fact that low-order Laplacian eigenmodes encode large-scale structure and global smoothness \cite{kumar2023unified, jin2020graph, loukas2018spectrally, dey2018spectralconcentrationgreedykclustering}. Retaining these modes helps preserve key partitioning and connectivity features that influence the solution space of the QUBO or Max-Cut objective. This aligns with prior work in spectral clustering and graph coarsening, where spectral energy correlates with the quality of preserved partitions and optimization potential.

\item \textit{Sensitivity to \(\alpha\).}  
The threshold \(\alpha\) controls the trade-off between compression and structural fidelity. A lower \(\alpha\) leads to more aggressive shrinking but may discard important structural components, potentially harming solution quality. Conversely, a higher \(\alpha\) retains more structure but limits reduction. In our experiments, we found solution quality to be relatively stable for \(\alpha \in [0.85, 0.95]\), with diminishing returns beyond \(\alpha = 0.95\). A more principled selection of \(\alpha\) could be problem-dependent, potentially guided by heuristics based on the graph spectrum or constraint density.

\item \textit{Interaction with Correlation-Based Shrinking.}  
The spectral method determines only the target node count \(k\); it does not guide which nodes are merged. The merging process itself is governed by correlation scores \(C_{ij}\), derived from the SDP relaxation of the Max-Cut QUBO. At each step, the pair of clusters with the strongest alignment or anti-alignment are merged, and correlations are updated accordingly.

Thus, spectral and correlation-based methods serve complementary roles: the spectral method provides a principled, global stopping condition, while the correlation-based mechanism guides the local refinement of the graph. This separation of concerns improves modularity and makes the framework adaptable to other stopping criteria (e.g., runtime or approximation gap).
\end{itemize}

\paragraph{Reconstruction:}

After solving the reduced problem, the solution for the original graph is reconstructed by backtracking through the recorded shrinking steps:
\begin{itemize}
    \item Reverse each shrinking step to determine the partitioning of the original nodes.
    \item Extend the solution iteratively until all nodes in the original graph are assigned to their respective partitions.
\end{itemize}

\section{Heuristic Enhancements to Preserve Feasibility}
\label{sec:new-addition}
This section presents two complementary heuristic strategies to enhance the feasibility and quality of solutions. We call them a \emph{proactive} strategy,  which incorporates constraint awareness directly into the structure of the heuristic itself and a \emph{reactive} strategy, which involves post-solution verification and repair.

\subsection{Proactive Strategy: Constraint-Aware Merging}

\subsubsection{General Idea}

Many combinatorial optimization problems benefit from graph shrinking strategies that reduce problem size via variable merging. Our proactive approach introduces constraint awareness into this merging process, thereby preventing infeasibility before it arises.

We employ a Semi-Definite Programming (SDP) relaxation, which produces a correlation matrix \( X \in \mathbb{R}^{n \times n} \), where \( X_{ij} = v_i^\top v_j \) reflects the geometric alignment of variables in a continuous embedding space.

A \emph{supernode} refers to a group of original problem variables that are merged and treated as a single unit during the graph shrinking process. Initially, each variable forms its own supernode, but as the heuristic progresses, variables or groups of variables that exhibit strong correlations are combined into larger supernodes. 

For supernodes \( C_i, C_j \subseteq V \), the standard merging strategy selects the pair that maximizes average correlation.

We refine the correlation matrix to now include an additional penalty term to discourage merges that risk violating constraints:

\[
S(C_i, C_j) = \mathbb{E}_{u \in C_i, v \in C_j} [X_{uv}] - \lambda \cdot \Pi(C_i, C_j)
\]

Here, $\mathbb{E}_{u \in C_i, v \in C_j} [X_{uv}]$ is the expected correlation, calculated as the arithmetic mean of the SDP-derived correlations over all pairs of the orginal nodes $(u,v)$ where $u \in C_i$ and $v \in C_j$, and \( \Pi(C_i, C_j) \) is a problem-specific penalty function and \( \lambda > 0 \) is a tunable hyperparameter that controls the strength of constraint penalization.

The value of \( \lambda \) is chosen using sensitivity analysis: we run the heuristic with varying \( \lambda \) and monitor the trade-off between solution quality and feasibility. An optimal \( \lambda \) balances aggressive merging with constraint satisfaction.

\subsubsection{Application to Multidimensional Knapsack Problem (MDKP)}

For MDKP, merging clusters \( C_i \) and \( C_j \) is risky if their combined weight is likely to exceed capacity in any dimension. The penalty is defined as:

\[
\Pi_{\text{MDKP}}(C_i, C_j) = \frac{1}{m} \sum_{k=1}^{m} \frac{\sum_{l \in C_i \cup C_j} W_{kl}}{C_k}
\]

This penalty captures the average normalized resource usage across all dimensions. Higher values indicate greater likelihood of violating capacity constraints, and thus discourage infeasible merges.

\subsubsection{Application to Maximum Independent Set (MIS)}

In MIS, merging supernodes that contain adjacent vertices results in infeasibility. We define:

\[
\Pi_{\text{MIS}}(C_i, C_j) = 
\begin{cases}
1 & \text{if } \exists u \in C_i, v \in C_j \text{ such that } (u,v) \in E \\
0 & \text{otherwise}
\end{cases}
\]

This ensures that supernodes connected by at least one edge are penalized maximally and discouraged from merging.

\subsubsection{Application to Quadratic Assignment Problem (QAP)}

For QAP, merging clusters \( C_u \) and \( C_v \) must respect the permutation structure of the solution. Each variable \( x_{ij} \) represents assigning facility \( i \) to location \( j \). To preserve feasibility, merges that could imply assigning the same facility to multiple locations or assigning multiple facilities to the same location are penalized.

The constraint-aware penalty is defined as:

\begin{equation}
\Pi_{\text{QAP}}(C_u, C_v) =
\begin{cases}
1, & \begin{aligned}
     & \exists\, x_{i_1 j_1} \in C_u,\ x_{i_2 j_2} \in C_v\ \text{such that} \\
     & i_1 = i_2\ \text{or}\ j_1 = j_2
     \end{aligned} \\
0, & \text{otherwise}
\end{cases}
\end{equation}

This penalty captures conflicts that would violate the one-to-one assignment requirement of QAP. A value of 1 indicates that the merge would couple variables corresponding to either the same facility or the same location, which is infeasible in any valid permutation. These merges are therefore strongly discouraged during the graph shrinking process.

\subsubsection{Discussion: Trade-offs and Design Rationale}

The integration of constraint-aware penalties into the graph shrinking process introduces a fundamental trade-off between global structure preservation and local constraint satisfaction. This is particularly evident in scenarios where the SDP relaxation suggests a strong merge—i.e., two variables \( i, j \) exhibit high correlation \( X_{ij} \)—but such a merge is discouraged due to a potential violation of original problem constraints (e.g., adjacency in MIS, or joint resource overload in MDKP).

This tension reflects a hybrid strategy: on one hand, the SDP matrix encodes global geometric insights about the solution landscape of the relaxed Max-Cut problem; on the other hand, the constraint-aware penalty imposes a local correction mechanism rooted in the structure of the original combinatorial problem. The penalized merge score,
\[
S(C_i, C_j) = \mathbb{E}_{u \in C_i, v \in C_j}[X_{uv}] - \lambda \cdot \Pi(C_i, C_j)
\]
thus serves as a balancing function that interpolates between these two perspectives.

Importantly, this mechanism may introduce a risk of over-penalization. A merge that appears infeasible when viewed locally (e.g., two high-weight items in MDKP) may, in fact, be critical to preserving a more subtle or non-local structure that leads to higher-quality solutions downstream. In this sense, the constraint-aware approach can act greedily, pruning globally valuable but locally risky merges.

To manage this trade-off, the penalty factor \( \lambda \) acts as a tunable parameter. A lower value of \( \lambda \) favors global correlation signals from the SDP and tolerates more risk, possibly increasing the burden on the subsequent repair mechanism. In contrast, a higher \( \lambda \) enforces strict local constraint adherence during the shrinking process, yielding more feasible intermediate solutions but potentially excluding complex merge patterns that contribute to better final outcomes.

Despite this limitation, constraint-aware shrinking remains a pragmatic heuristic. It does not aim to guarantee global optimality—which remains infeasible in many of the targeted problem classes—but instead seeks to bias the search toward more promising, constraint-respecting regions of the solution space. Empirical results support the efficacy of this design: across tested instances, the method improves pre-repair feasibility rates while maintaining post-repair solution quality, suggesting that the proactive bias introduced by constraint-aware merging is beneficial on balance.

\subsection{Reactive Strategy: Verification and Repair}

\subsubsection{General Idea}

Let an optimization problem be defined as \( \mathcal{P} = (f, \mathcal{C}) \), where \( f: \{0,1\}^n \rightarrow \mathbb{R} \) is the objective function to be maximized, and \( \mathcal{C} \) is a set of hard constraints. A candidate solution is a binary vector \( x \in \{0, 1\}^n \).

Given a heuristic algorithm \( \mathcal{H} \), the solution \( x' = \mathcal{H}(\mathcal{P}) \) is not guaranteed to be feasible. To address this, we apply a two-stage \textit{reactive correction} procedure:

\begin{itemize}
  \item \textbf{Verification.} A boolean function \( \mathcal{V}(x', \mathcal{P}) \) evaluates whether all constraints in \( \mathcal{C} \) are satisfied:
  \[
  \mathcal{V}(x', \mathcal{P}) = 
  \begin{cases}
    \texttt{True}, & \text{if } x' \text{ satisfies all constraints}, \\
    \texttt{False}, & \text{otherwise}.
  \end{cases}
  \]

  \item \textbf{Repair.} If \( \mathcal{V}(x', \mathcal{P}) = \texttt{False} \), a greedy repair operator \( \mathcal{R} \) is invoked to obtain a new solution \( x'' = \mathcal{R}(x', \mathcal{P}) \). This operator removes the minimal number of elements necessary to restore feasibility while seeking to minimize degradation of \( f(x'') \).
\end{itemize}

The greedy repair strategy proceeds iteratively: it identifies the smallest local modification (e.g., removing one variable from the solution) that leads to the greatest improvement in constraint satisfaction. The removed variable is selected using a domain-specific heuristic, such as impact on objective or involvement in constraint violations. The process repeats until a feasible solution is obtained or no further beneficial actions are available.

\subsubsection{Application to Multidimensional Knapsack Problem (MDKP)}

Let \( n \) be the number of items and \( m \) the number of resource constraints. Each item has a profit \( p_i \) and weight \( W_{ji} \) in dimension \( j \). The goal is to maximize profit while satisfying capacity constraints.

\begin{itemize}
  \item \textbf{Objective:} \( f(x) = p^\top x \)
  \item \textbf{Constraint:} \( \sum_{i=1}^{n} W_{ji} x_i \leq C_j \quad \forall j \in \{1, \dots, m\} \)
\end{itemize}

\paragraph{Verification.} Compute \( Wx \leq C \), a component-wise inequality, where \( W \in \mathbb{R}^{m \times n} \), \( x \in \{0,1\}^n \), and \( C \in \mathbb{R}^m \).

\paragraph{Repair.} The \texttt{MDKPGreedyRepair} operator proceeds as follows:

\begin{enumerate}
  \item Identify violated dimensions \( D_v = \{j \mid (Wx)_j > C_j\} \)
  \item Select the most violated dimension \( j^* \)
  \item For all \( i \) with \( x_i = 1 \), compute efficiency ratios \( \rho_i = p_i / W_{j^* i} \)
  \item Remove the item with minimum \( \rho_i \): \( k = \arg\min \rho_i \), by setting \( x_k = 0 \)
  \item Repeat until \( D_v = \emptyset \)
\end{enumerate}

\subsubsection{Application to Maximum Independent Set (MIS)}

Given a graph \( G = (V, E) \) with \( |V| = n \), an independent set is a subset \( S \subseteq V \) such that no two vertices in \( S \) are adjacent. A solution is encoded as \( x \in \{0, 1\}^n \), where \( x_i = 1 \) implies \( i \in S \).

\begin{itemize}
  \item \textbf{Objective:} \( f(x) = \sum_{i=1}^{n} x_i \)
  \item \textbf{Constraint:} \( x_i + x_j \leq 1 \quad \forall (i, j) \in E \)
\end{itemize}

\paragraph{Verification.} The solution \( x \) is feasible if for all \( i, j \in V \), \( x_i A_{ij} x_j = 0 \), where \( A \) is the adjacency matrix.

\paragraph{Repair.} The \texttt{MISGreedyRepair} operator removes one vertex from each edge in the conflict set \( E_c = \{(i, j) \in E \mid x_i = x_j = 1\} \). The vertex with higher degree is removed:
\[
k = \arg\max_{i \in \{u, v\}} \deg(i)
\]
This aims to preserve the less-connected (peripheral) nodes. The process repeats until \( E_c = \emptyset \).

\subsubsection{Application to Quadratic Assignment Problem (QAP)}

Let \( n \) be the number of facilities and locations. The goal is to assign each facility to exactly one location such that the total cost—computed as the sum of flow between facilities times the distance between assigned locations—is minimized.

\begin{itemize}
  \item \textbf{Objective:} \( f(x) = \sum_{i,k=1}^n \sum_{j,l=1}^n F_{ik} D_{jl} x_{ij} x_{kl} \)
  \item \textbf{Constraints:}
  \begin{itemize}
    \item Each facility is assigned to exactly one location: \( \sum_{j=1}^n x_{ij} = 1 \quad \forall i \in \{1, \dots, n\} \)
    \item Each location receives exactly one facility: \( \sum_{i=1}^n x_{ij} = 1 \quad \forall j \in \{1, \dots, n\} \)
  \end{itemize}
\end{itemize}

\paragraph{Verification.} Given a bitstring solution flattened into \( x \in \{0,1\}^{n^2} \), reshape into an \( n \times n \) matrix \( X \). Verify two conditions:
\begin{itemize}
  \item Every facility is assigned to one location: each row of \( X \) sums to 1.
  \item Every location receives one facility: each column of \( X \) sums to 1.
\end{itemize}

\paragraph{Repair.} The \texttt{repair\_qap\_solution} operator uses a linear assignment approach:

\begin{enumerate}
  \item Construct an \( n \times n \) cost matrix \( C \), where \( C_{ij} = -1 \) if \( x_{ij} = 1 \), and \( 0 \) otherwise.
  \item Solve the linear assignment problem \( \min_{\pi} \sum_{i=1}^n C_{i\pi(i)} \) using the Hungarian algorithm to find the best one-to-one assignment.
  \item The result is a valid permutation \( \pi \) mapping facilities to locations that maximally agrees with the "votes" in the original bitstring.
  \item Construct the final feasible solution matrix from \( \{(i, \pi(i)) \} \).
\end{enumerate}

This repair mechanism ensures feasibility by enforcing the permutation constraint while retaining the most confident suggestions from the QUBO output.

\section{\label{sec:level3} Quantum Optimization Framework}

The graph shrinking methodology presented in Algorithm~\ref{alg:graph_shrinking_reactive_proactive} provides a systematic approach to reducing the dimensionality of complex combinatorial optimization problems while preserving their essential structural properties. This method is particularly well-suited for integration with quantum optimization techniques, such as the Quantum Approximate Optimization Algorithm (QAOA) and the Variational Quantum Eigensolver (VQE). By reducing problem size, graph shrinking allows quantum algorithms to operate within the constraints of current hardware while maintaining the fidelity of the reconstructed solution to the original problem.

\subsection{Motivation for Combining Graph Shrinking and Quantum Techniques}

The exponential growth of computational complexity in large-scale combinatorial optimization problems presents significant challenges for classical methods. Quantum optimization algorithms, such as the Quantum Approximate Optimization Algorithm (QAOA) and the Variational Quantum Eigensolver (VQE), are designed to solve quadratic unconstrained binary optimization (QUBO) problems more efficiently. However, these approaches are inherently limited by the number of qubits available on current quantum hardware. The graph shrinking process mitigates this constraint by reducing the problem’s dimensionality, allowing quantum algorithms to address instances that would otherwise exceed hardware capabilities.

\subsection{Proposed Workflow}

The proposed hybrid classical--quantum workflow integrates constraint-aware graph shrinking with quantum optimization in a principled manner. The workflow comprises four key stages:

\begin{enumerate}
    \item \textbf{QUBO $\rightarrow$ Weighted Max-Cut:}  
    The original combinatorial optimization problem is first reformulated as a QUBO (Quadratic Unconstrained Binary Optimization) problem. Leveraging known equivalence transformations~\cite{barahona1989experiments}, the QUBO is then mapped to a Weighted Max-Cut problem, enabling structure-aware preprocessing and embedding techniques.

    \item \textbf{Proactive Graph Shrinking (Constraint-Aware):}  
    The graph is iteratively reduced from the original \( G \) to a compressed graph \( G' \) with \( k \) nodes using a constraint-aware merging heuristic. Guided by an SDP relaxation, the method selects variable pairs with strong geometric correlation while penalizing merges that are likely to violate original problem constraints. A tunable penalty factor \( \lambda \) balances global correlation strength against local constraint risk. All merging operations are recorded to enable downstream solution reconstruction.

    \item \textbf{Quantum Optimization on Reduced Problem:}  
    The compressed graph \( G' \) is encoded back into a reduced QUBO instance and solved using a quantum algorithm such as QAOA or CVaR-VQE. Operating in the smaller solution space enhances scalability and enables deeper quantum circuit execution within hardware constraints.

    \item \textbf{Reactive Solution Reconstruction and Repair:}  
    The quantum solution on \( G' \) is lifted back to the original variable space via reverse replay of the recorded merging steps. If the reconstructed solution violates any of the original problem’s constraints, a problem-specific greedy repair heuristic is invoked. This ensures feasibility while preserving the structural advantages introduced by quantum optimization.
\end{enumerate}

\subsection{Implementation Considerations}

Key factors influencing the implementation of this approach include:
\begin{itemize}
    \item \textbf{Selection of Quantum Algorithm:}
    \begin{itemize}
        \item \textbf{QAOA} is well-suited for solving combinatorial optimization problems by iteratively refining solution quality. Its performance depends on the depth of the ansatz (\textit{p}), which directly affects circuit complexity and execution time.
        \item \textbf{VQE} is particularly effective for problems with complex objective functions, especially those formulated as Hamiltonians. It is often preferred for near-term quantum devices due to its hybrid quantum-classical nature, which mitigates hardware limitations.
    \end{itemize}
    The selection of a quantum algorithm depends on both the problem structure and the available quantum hardware. Any QUBO-solving quantum method, such as variants of \textbf{QAOA}~\cite{Blekos_2024} or the \textbf{QRAO} method~\cite{fuller2024approximate}, can be integrated within this framework. The choice of algorithm involves balancing factors such as qubit requirements, circuit depth, and noise resilience, with pre-processing techniques like graph shrinking playing a crucial role in optimizing resource utilization.

    \item \textbf{Accuracy in Solution Reconstruction:} 
The fidelity of the reconstructed solution is highly dependent on the accuracy of the correlations computed during the graph shrinking phase. Any inaccuracies in correlation estimation can propagate through the reduction process, potentially affecting the quality of the final solution. To mitigate this, regular recalculation of correlations can be performed at each iteration, ensuring robustness and stability in reconstruction. 

Additionally, adaptive thresholding techniques can be employed to dynamically refine correlation values, reducing sensitivity to noise and preserving the structural integrity of the original problem. The use of spectral methods in adaptive graph shrinking further enhances accuracy by leveraging global structural properties rather than local correlations alone.

    \item \textbf{Hardware Constraints:} The target size \( k \) for the reduced graph must align with the number of qubits available on the selected quantum hardware. Overly aggressive reduction can compromise solution quality.
\end{itemize}

\subsection{Bridging the Classical-Quantum Divide}

The graph shrinking methodology serves as a bridge between classical and quantum optimization paradigms, addressing the limitations of both. By leveraging classical reductions to enable quantum feasibility and utilizing quantum methods for high-quality solutions, this hybrid approach charts a practical pathway for solving large-scale optimization problems. Its versatility and scalability hold promise for advancing the frontier of combinatorial optimization across diverse domains.

\section{\label{sec:level7}Experimental Setup and Results}

\subsection{Environment}

The experiments were conducted on a high-performance server equipped with an Intel(R) Xeon(R) Gold 6154 CPU @ 3.00GHz, featuring 144 CPUs across four sockets, with 18 cores per socket and two threads per core. Quantum computations were simulated using the \texttt{Qiskit AerSimulator} with the matrix product state method, while classical preprocessing and graph reduction were performed using Python libraries such as \texttt{NetworkX} and \texttt{Docplex}. The graph shrinking algorithm was applied to reduce the size of classically challenging benchmark instances of the MDKP, MIS and QAP.

Our workflow begins with the formulation of problem instances as linear programs (LPs), which are then systematically encoded as Quadratic Unconstrained Binary Optimization (QUBO) problems via explicit constraint penalization. These QUBO formulations are subsequently transformed into equivalent weighted Max-Cut instances through the introduction of a single auxiliary node. 

We then apply our graph shrinking algorithm to the weighted Max-Cut representation, iteratively reducing the problem size while preserving critical constraint-encoded structure via correlation-aware and constraint-aware merging strategies. The reduced instances are solved using quantum optimization methods (VQE), executed on a quantum simulator. Finally, the resulting solutions are mapped back from the Max-Cut representation to the original QUBO form, and subsequently decoded into valid solutions for the original LP formulation.

\subsection{Metrics}

The performance of the proposed approach was evaluated using the following metrics:
\begin{itemize}
    \item \textbf{Optimality Gap (\%)}: The percentage difference between the obtained solution and the known optimal solution.
\begin{align}
\text{Opt. Gap (\%)} = 
    & \frac{
    \text{Obj. Best} 
    - \text{Obj. Obtained}
    }{
    \text{Obj. Best}
    } \nonumber 
    & \times 100
\end{align}

    \item \textbf{Relative Solution Quality (\%):} This metric evaluates the quality of the obtained solution relative to the best-known or optimal solution. It is typically expressed as a percentage, calculated as:
\begin{equation}
\text{RSQ (\%)} = \left( 
    \frac{\text{Obj. Value of Obtained Solution}}
         {\text{Obj. Value of Best-Known Solution}} 
\right) \times 100
\end{equation}
Higher values indicate solutions closer to the optimal, demonstrating the effectiveness of the algorithm or approach used.

    \item \textbf{Size}: The number of variables (qubits) in the reduced problem after graph shrinking.
\end{itemize}

\subsection{Sensitivity Analysis of the Penalty Factor}

We conducted a sensitivity analysis to study the effect of the penalty factor \( \lambda \) on solution quality. This factor governs the strength of constraint-awareness during the graph shrinking phase by penalizing merges that are likely to violate original problem constraints.

The analysis was performed on randomly sampled instances from both the Maximum Independent Set (MIS) and Multidimensional Knapsack Problem (MDKP) classes. For each instance, we executed the full pipeline across a range of values \( \lambda \in \{0.0, 0.5, 1.0, 1.5, 2.0, 5.0, 10.0\} \).

We observed that low values of \( \lambda \) (e.g., 0.0–1.0) encouraged aggressive merges, often resulting in infeasible solutions that required substantial post-repair. In contrast, moderate values (around 1.0–2.0) achieved a more favorable balance between structural exploration and constraint satisfaction. Very high values (e.g., \( \lambda = 10.0 \)) tended to overemphasize feasibility, occasionally missing globally high-quality configurations.

Overall, setting \( \lambda \) within the range of 1.0 to 2.0 was found to be robust across problem types, enabling the proactive avoidance of infeasibility while preserving solution quality.

\subsection{Results}

We evaluated our graph shrinking framework on three benchmark problem classes: \textbf{Multi-Dimensional Knapsack Problem (MDKP)}, \textbf{Maximum Independent Set (MIS)}, and \textbf{Quadratic Assignment Problem (QAP)}. For each instance, we applied adaptive graph shrinking to reduce problem size and then solved the reduced problem using classical (CPLEX) and quantum (VQE) solvers. For all CPLEX-based experiments, we imposed a time limit of 14,400 seconds (4 hours) to ensure consistency across large-scale instances and fair comparison with quantum solvers.

To assess the impact of shrinking extent, we conducted an ablation study comparing fixed-ratio reductions (\(\lfloor 2/3 \cdot n \rfloor\), \(\lfloor 1/2 \cdot n \rfloor\)) with adaptive shrinking. Results (Appendix~\ref{app:shrinking_abalation}) show that while aggressive shrinking reduces runtime, it can increase the optimality gap. Adaptive shrinking provides the best trade-off, balancing solution quality and resource efficiency. 

\begin{figure*}[htbp]
    \centering
    \includegraphics[width=0.9\textwidth]{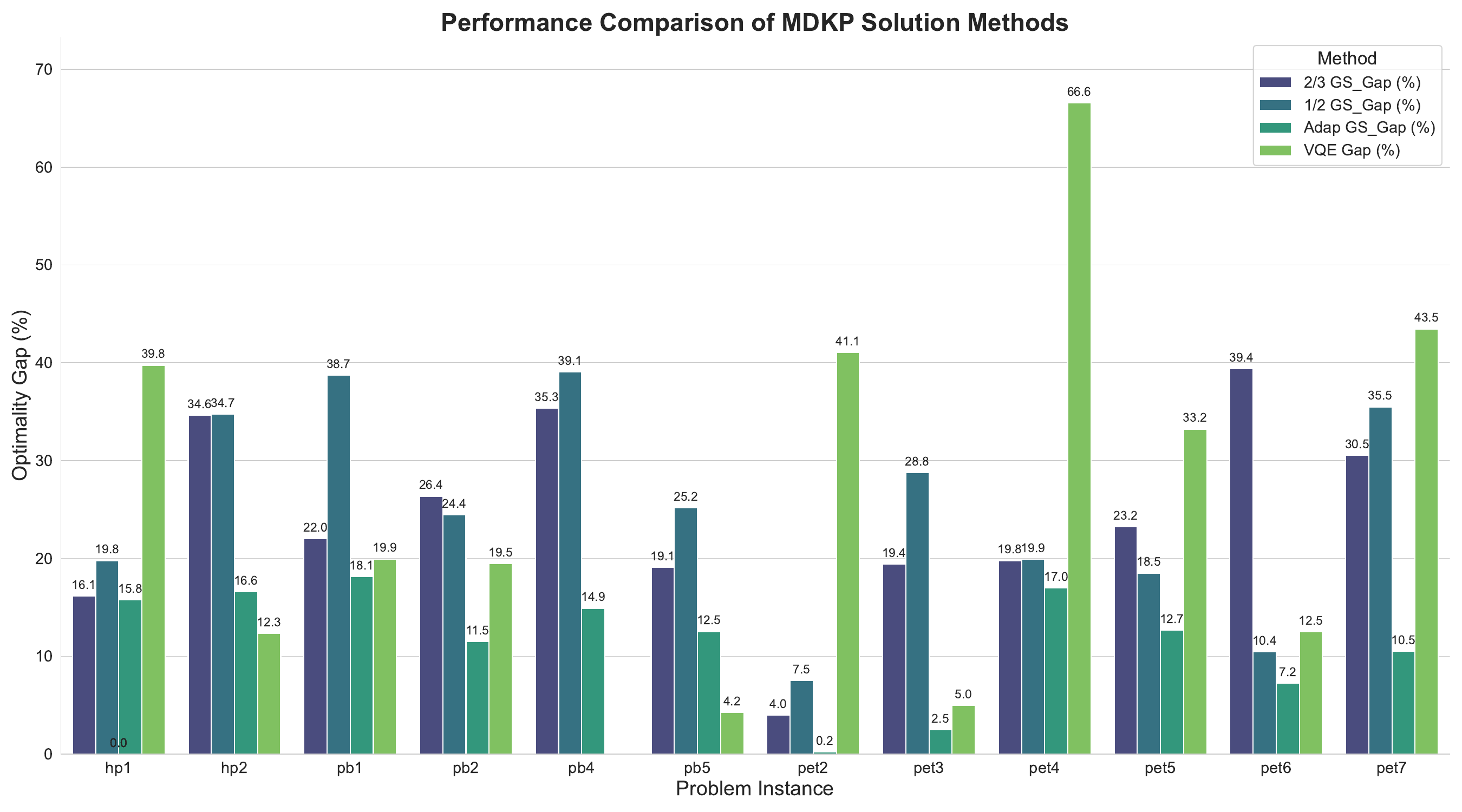}
    \caption{\label{fig:mdkp_shrink_plot}Optimality gap (\%) for various shrinking strategies on MDKP benchmark instances. Fixed strategies reduce problem size to $\lfloor \frac{2}{3}n \rfloor$ or $\lfloor \frac{1}{2}n \rfloor$, while Adaptive Shrinking dynamically adjusts based on instance structure. Lower gap values indicate better solution quality. Adaptive Shrinking consistently performs better, except a few cases.}
\end{figure*}

Figure~\ref{fig:mdkp_shrink_plot} illustrates the optimality gap across different shrinking strategies on MDKP instances, highlighting the trade-offs between fixed-ratio reductions and adaptive approaches.

For quantum solving, we tested both constraint-aware and non-constraint-aware QUBO formulations, while CPLEX was applied to the non-constraint-aware version. Our experiments reveal that:

\begin{itemize}
    \item \textbf{Adaptive shrinking} significantly reduces problem size while preserving solution quality, enabling quantum solvers to handle otherwise intractable instances.
    \item \textbf{VQE with constraint-aware formulations} generally yields better solution quality (higher RSQ or lower optimality gap) than the non-aware version, demonstrating the value of embedding constraints in the QUBO.
    \item \textbf{CPLEX on shrunken instances} performs competitively and provides a strong classical baseline, even without explicit constraint modeling.
\end{itemize}

Performance trends are summarized in the following subsections, which show optimality gaps, runtimes, perofrmance and time phase breakdown across solvers and formulations. Complete tabular results are provided in Appendix~\ref{app:table}.

\subsection*{Comparison of Solver Performance based on Solution Quality}

\begin{figure*}[htbp]
    \centering
    \includegraphics[width=\textwidth]{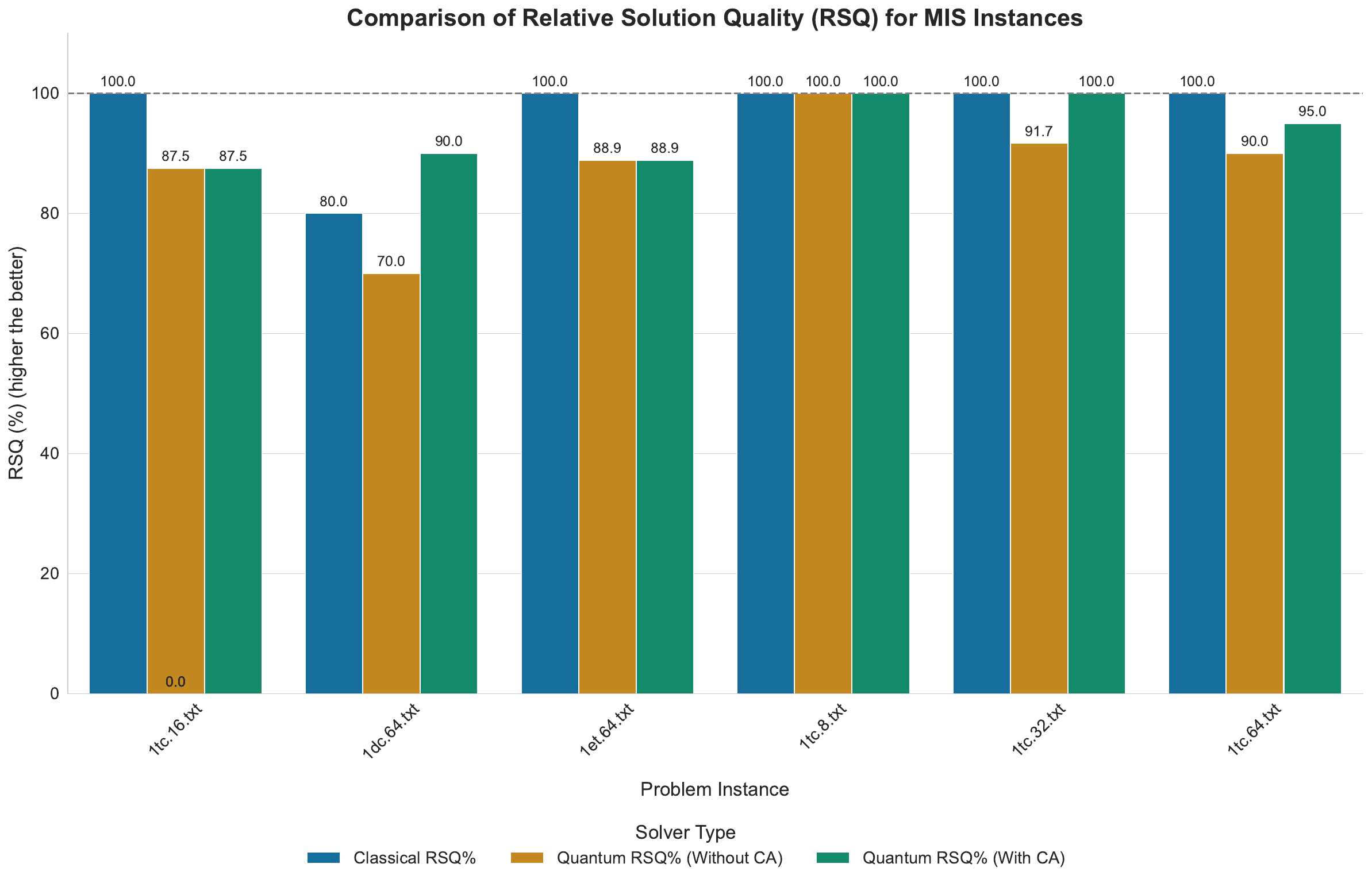}
    \caption{\label{fig:mis_rsq_comparison}Comparison of Relative Solution Quality (RSQ\%) for MIS benchmark instances. RSQ is defined as the ratio of the obtained solution size to the known optimal value. Classical CPLEX results serve as the baseline, while quantum results are shown for VQE with and without constraint-aware QUBO formulations. Higher values are better.}
\end{figure*}

\begin{figure*}[htbp]
    \centering
    \includegraphics[width=\textwidth]{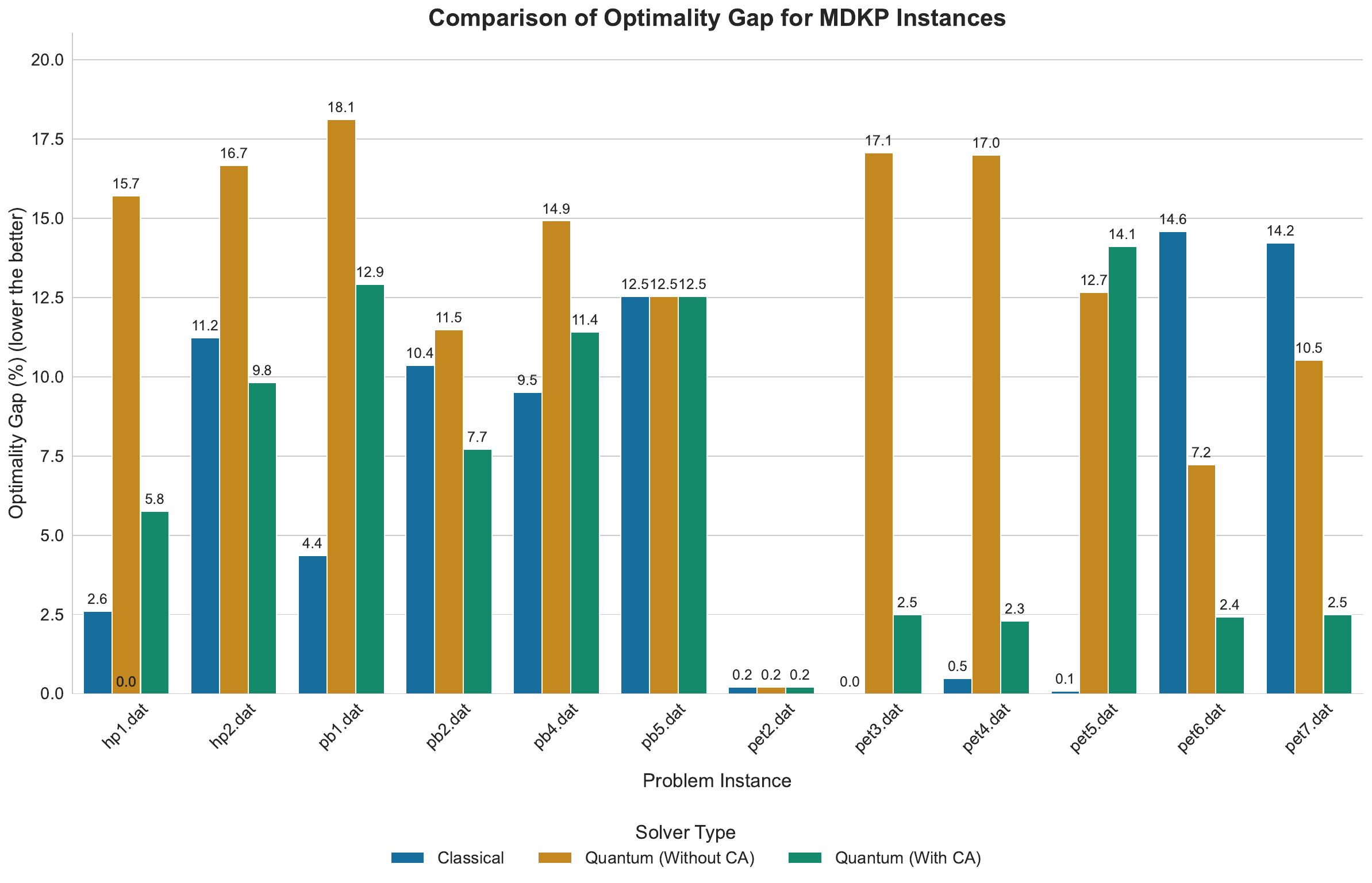}
    \caption{\label{fig:mdkp_gap_comparison}Comparison of Optimality Gap (\%) for MDKP benchmark instances. The optimality gap measures the percentage deviation from the known optimal objective value. Lower values are better. Classical CPLEX and quantum VQE results are shown for both non-constraint-aware and constraint-aware QUBO formulations.}
\end{figure*}

\begin{figure*}[htbp]
    \centering
    \includegraphics[width=\textwidth]{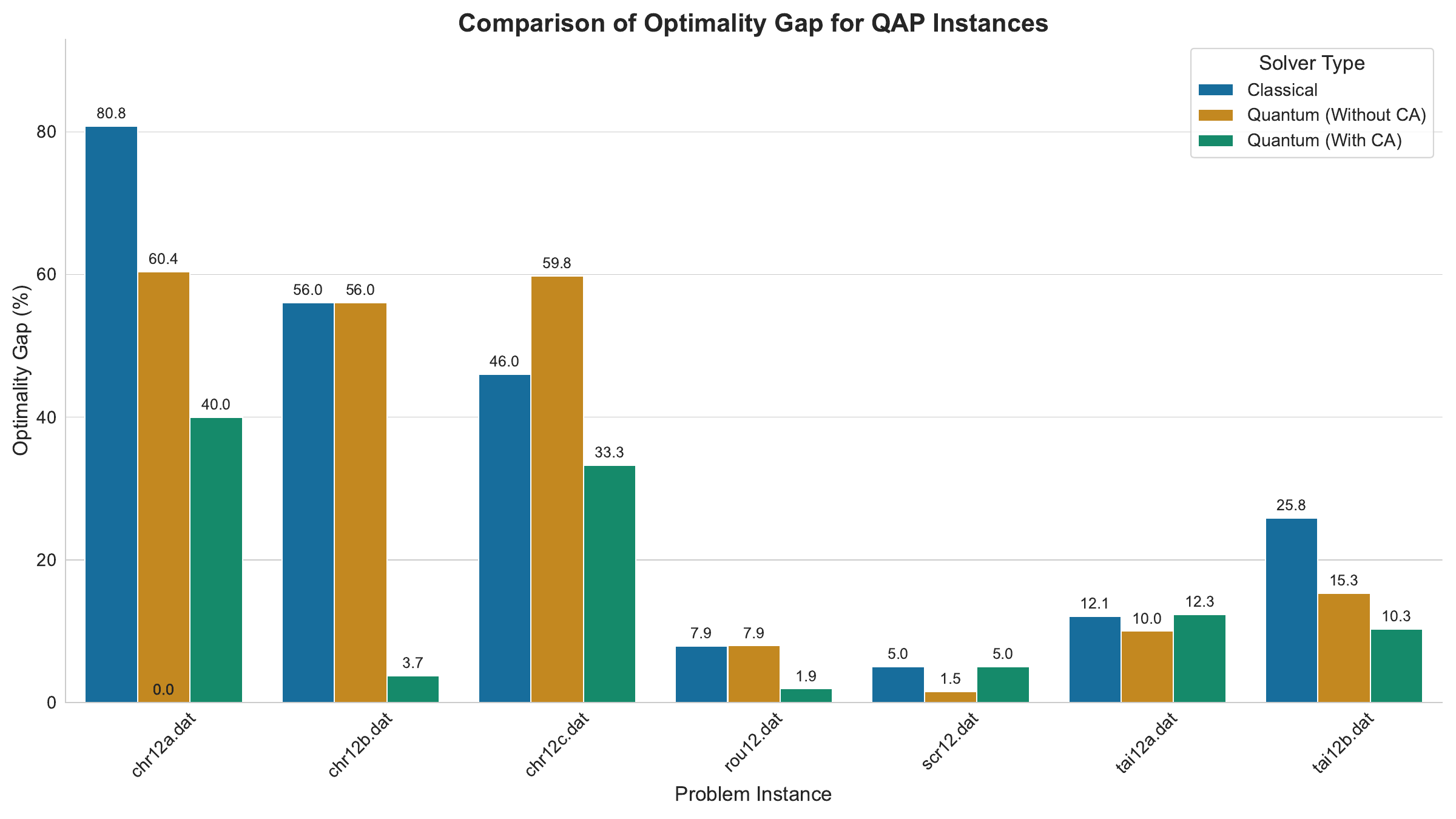}
    \caption{\label{fig:qap_gap_comparison}Comparison of Optimality Gap (\%) for QAP benchmark instances. The optimality gap measures the percentage deviation from the known optimal objective value. Lower values are better. Classical CPLEX and quantum VQE results are shown for both non-constraint-aware and constraint-aware QUBO formulations.}
\end{figure*}

Figure~\ref{fig:mis_rsq_comparison}, Figure~\ref{fig:mdkp_gap_comparison}, and Figure~\ref{fig:qap_gap_comparison} present a comparative evaluation of solver performance on the Maximum Independent Set (MIS), Multi-Dimensional Knapsack Problem (MDKP), and Quadratic Assignment Problem (QAP) instances, respectively.

For MIS, we report the \textbf{Relative Solution Quality (RSQ\%)}, defined as the percentage ratio of the obtained solution size to the known optimal value. For MDKP and QAP, we report the \textbf{Optimality Gap (\%)}, which measures the deviation of the obtained solution from the known optimal objective, with lower values indicating better performance.

In all three cases, we compare three solver variants:
\begin{itemize}
    \item Classical CPLEX applied to the adaptively shrunken problem (baseline).
    \item VQE applied to the non-constraint-aware QUBO formulation.
    \item VQE applied to the constraint-aware QUBO formulation.
\end{itemize}

As shown in the figures, CPLEX consistently yields high-quality solutions across all instances. Notably, VQE performance improves significantly when constraints are embedded in the QUBO, underscoring the effectiveness of constraint-aware formulations. These results highlight the combined value of adaptive shrinking and constraint modeling in improving quantum optimization outcomes.

\subsection*{Comparison of total runtime based on problem size}

\begin{figure*}[htbp]
    \centering
    \includegraphics[width=0.9\textwidth]{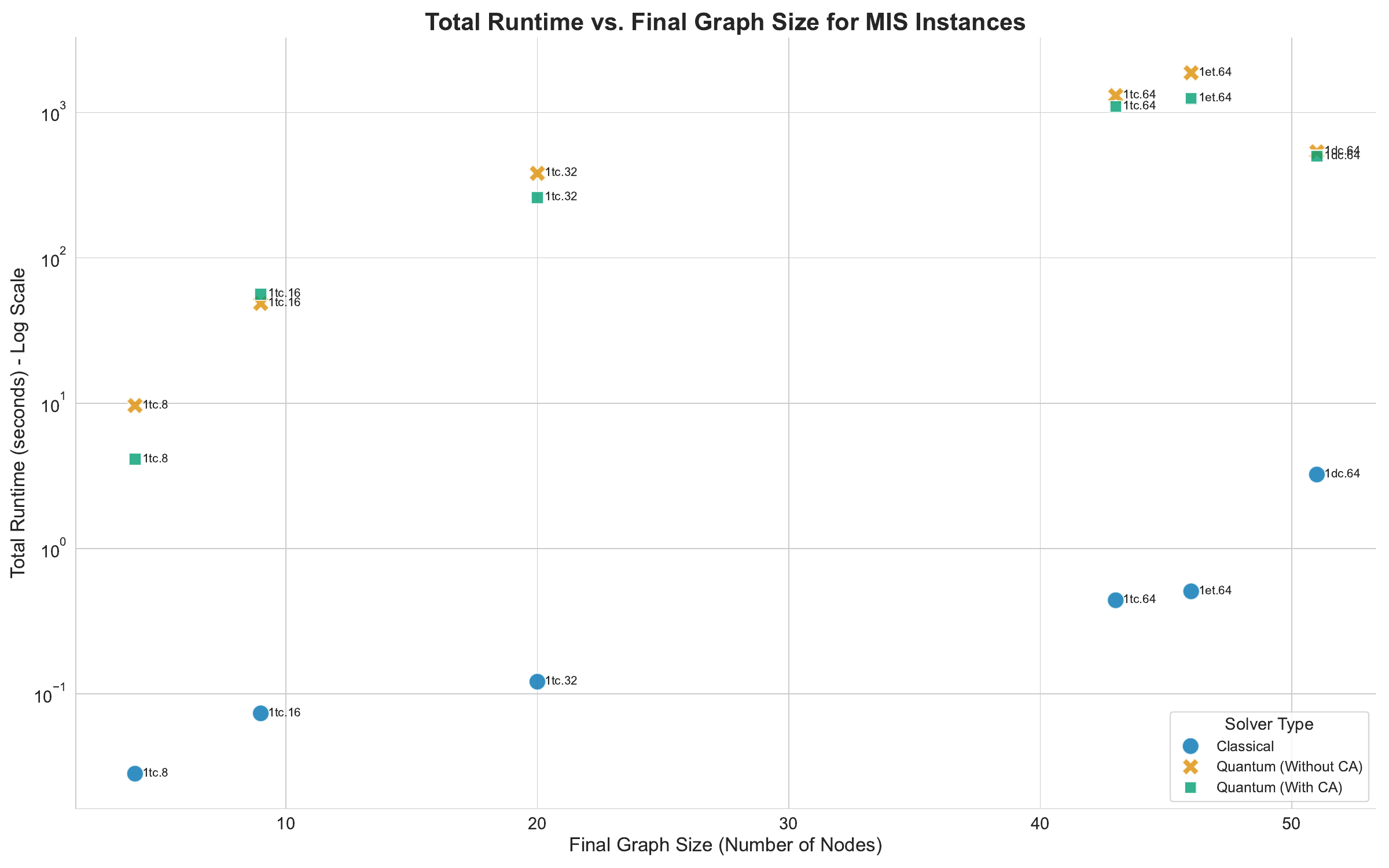}
    \caption{\label{fig:mis_runtime_size}Total runtime (log scale) vs. final graph size for MIS instances. Classical solvers are consistently efficient. Quantum runtimes are way higher than the classical counterparts.}
\end{figure*}

\begin{figure*}[htbp]
    \centering
    \includegraphics[width=0.9\textwidth]{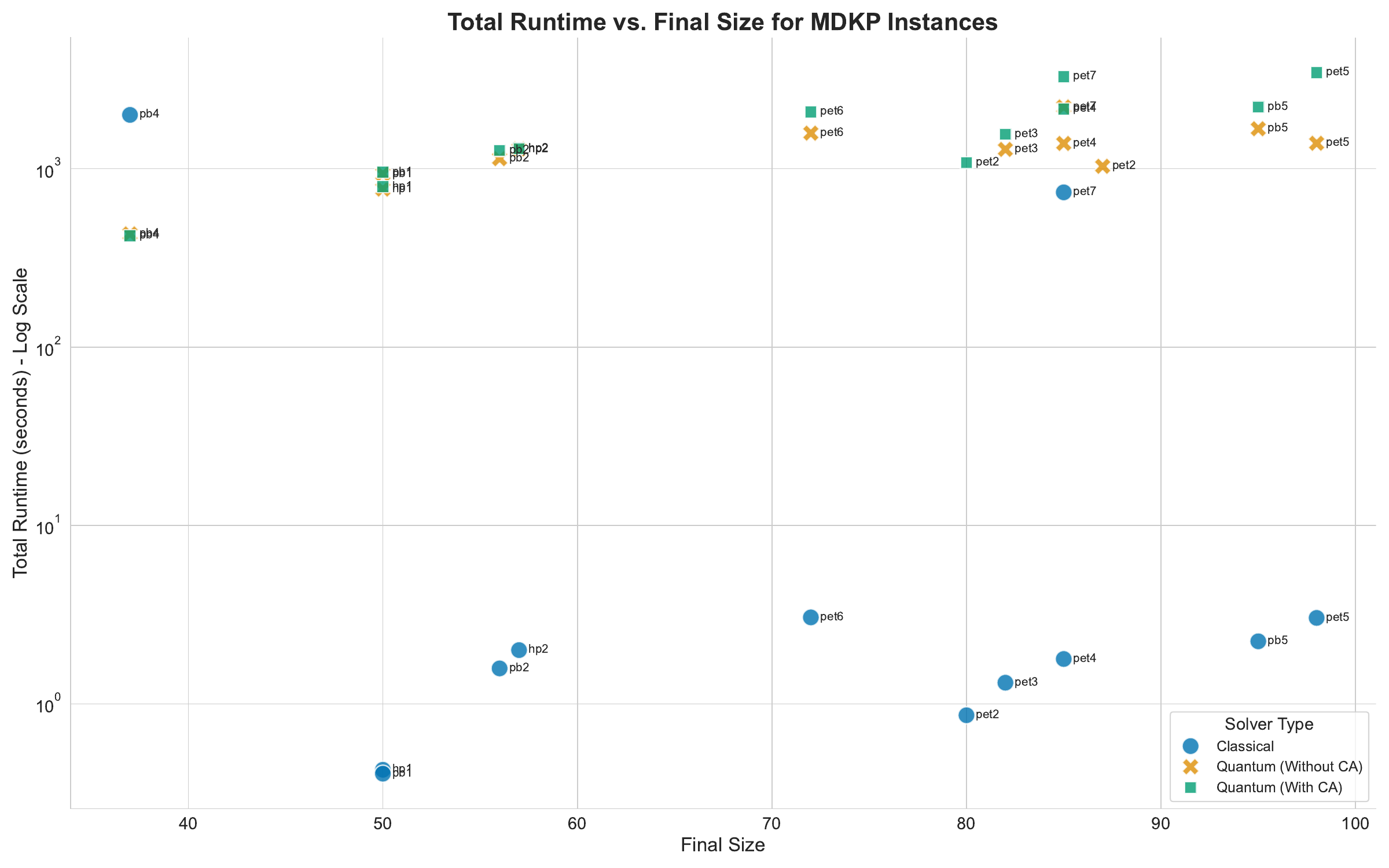}
    \caption{\label{fig:mdkp_runtime_size}Total runtime (log scale) vs. final solution size for MDKP instances. Classical solvers dominate in efficiency, while constraint-aware shrinking increases quantum runtimes.}
\end{figure*}

\begin{figure*}[htbp]
    \centering
    \includegraphics[width=0.9\textwidth]{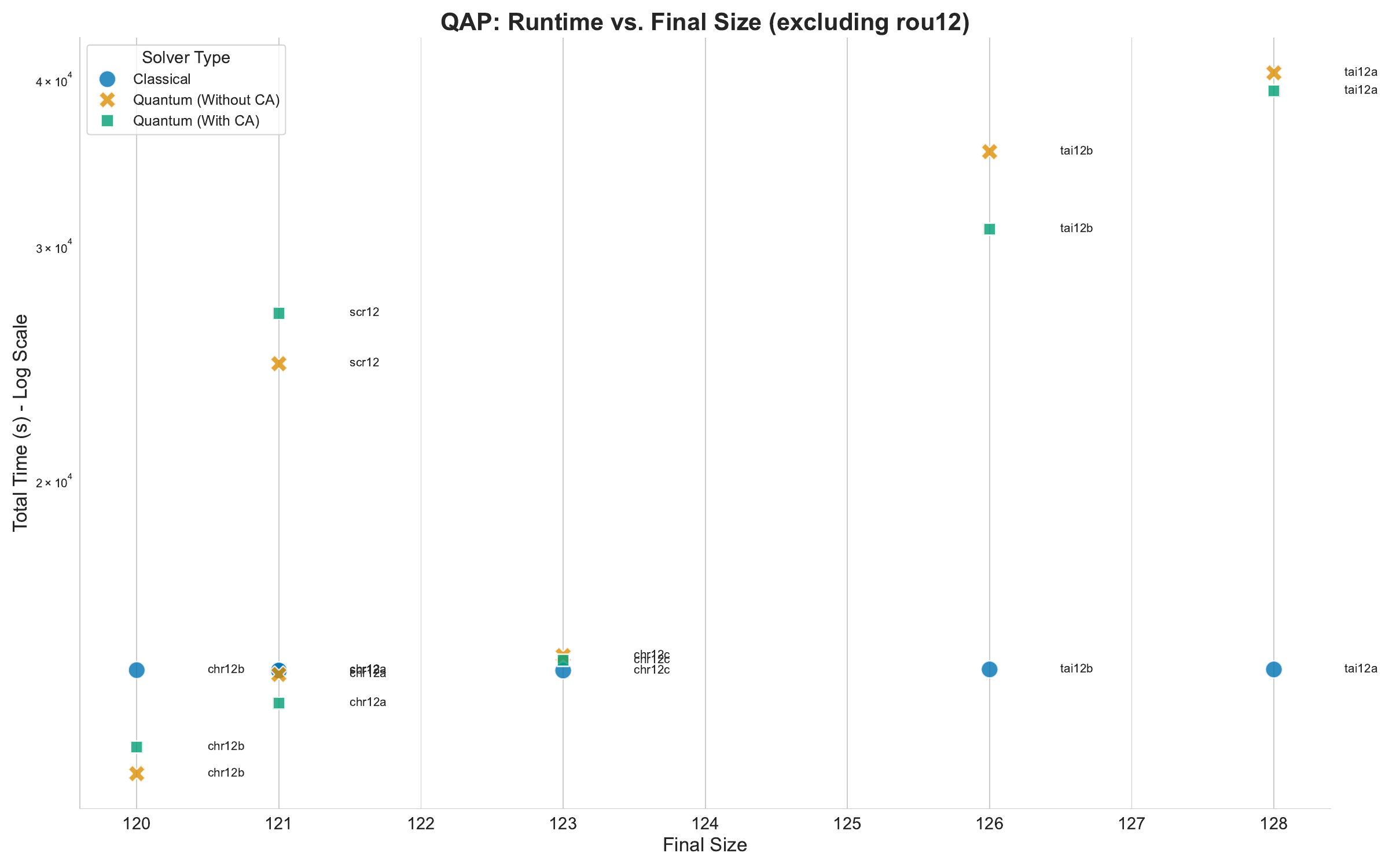}
    \caption{\label{fig:qap_runtime_size}Total runtime (log scale) vs. final solution size for QAP instances. Runtimes remain high due to problem difficulty, but constraint-aware shrinking enhances quantum efficiency.}
\end{figure*}

\vspace{1em}
\noindent
Beyond solution quality, we further analyze the \textbf{total runtime} of each solver with respect to the \textbf{final problem size} after graph shrinking. Figure~\ref{fig:mis_runtime_size}, Figure~\ref{fig:mdkp_runtime_size}, and Figure~\ref{fig:qap_runtime_size} present the runtime (in seconds, log scale) as a function of the reduced instance size for MIS, MDKP, and QAP benchmark families, respectively.

\medskip
\noindent
These plots offer additional insights into the runtime scalability of the solver variants:
\begin{itemize}
    \item For \textbf{MIS}, classical CPLEX is extremely fast, particularly due to the substantial reduction in graph size. Quantum solvers exhibit increasing runtime with instance size, but instances derived from \emph{constraint-aware shrinking} result in more efficient execution than those generated without constraint awareness.
    
    \item For \textbf{MDKP}, classical solvers are significantly faster across the board. While VQE incurs higher computational cost, instances obtained through no constraint-awared shrinking lead to consistently shorter runtimes compared to their constrained counterparts.
    
    \item For \textbf{QAP}, the runtime is generally high due to problem complexity. Nevertheless, even in this challenging setting, constraint-aware shrinking yields tangible runtime improvements for quantum solvers.
\end{itemize}

\noindent

Overall, these results confirm that constraint-aware shrinking improves solution quality across all benchmarks, although it incurs slightly higher runtimes due to the additional computation involved in evaluating penalized correlations. This trade-off highlights the value of integrating constraint information into the shrinking process to obtain more structured and feasible subproblems for quantum optimization.

\subsection*{Computational Resources Phase Breakdown}

\begin{figure*}[htbp]
    \centering
    \includegraphics[width=0.9\textwidth]{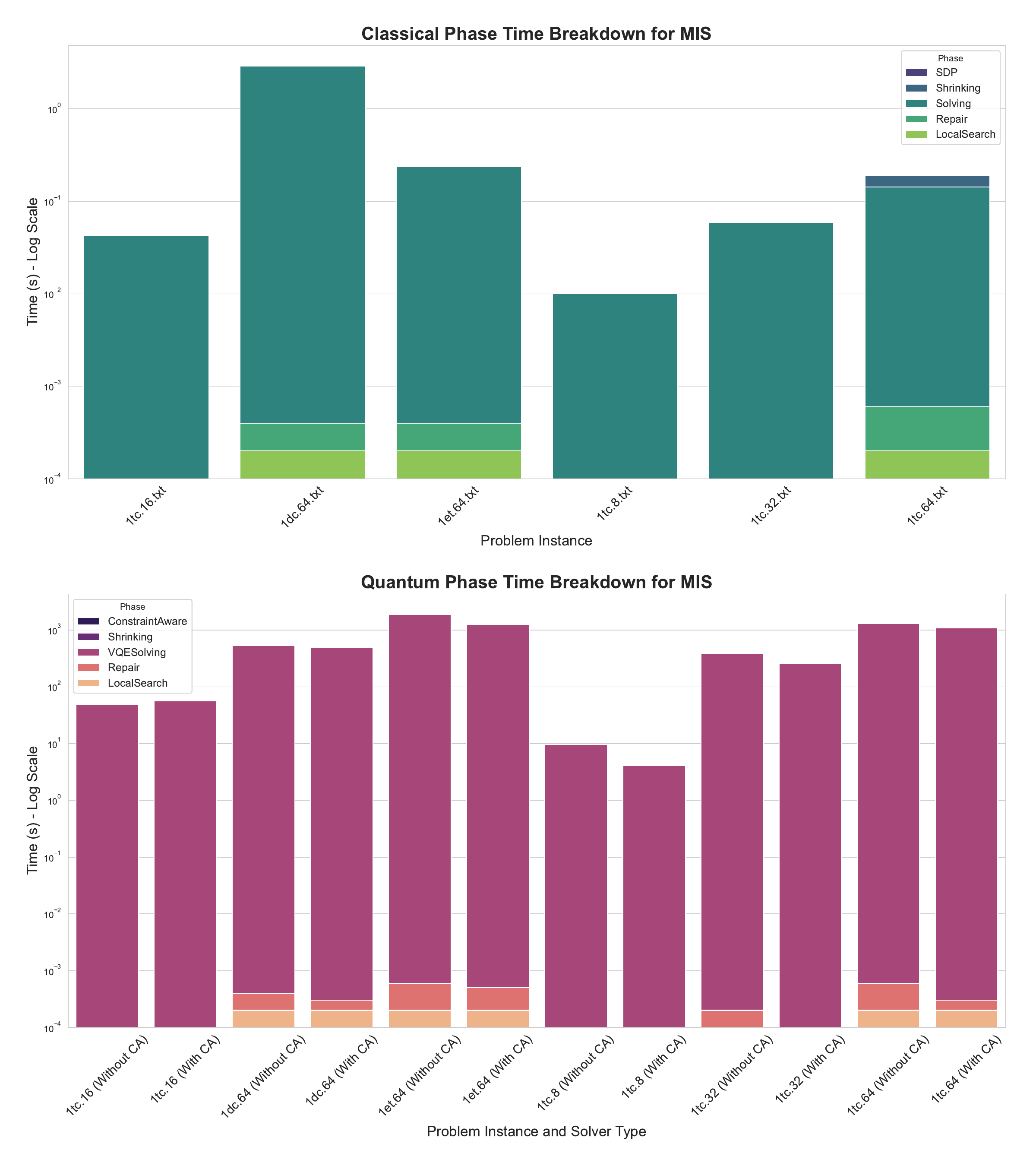}
    \caption{\label{fig:mis_phase_breakdown}Phase-wise runtime breakdown (log scale) for MIS instances. Each bar represents an instance with segments for shrinking, QUBO conversion, and solving. Classical and quantum pipelines are compared.}
\end{figure*}

\begin{figure*}[htbp]
    \centering
    \includegraphics[width=0.9\textwidth]{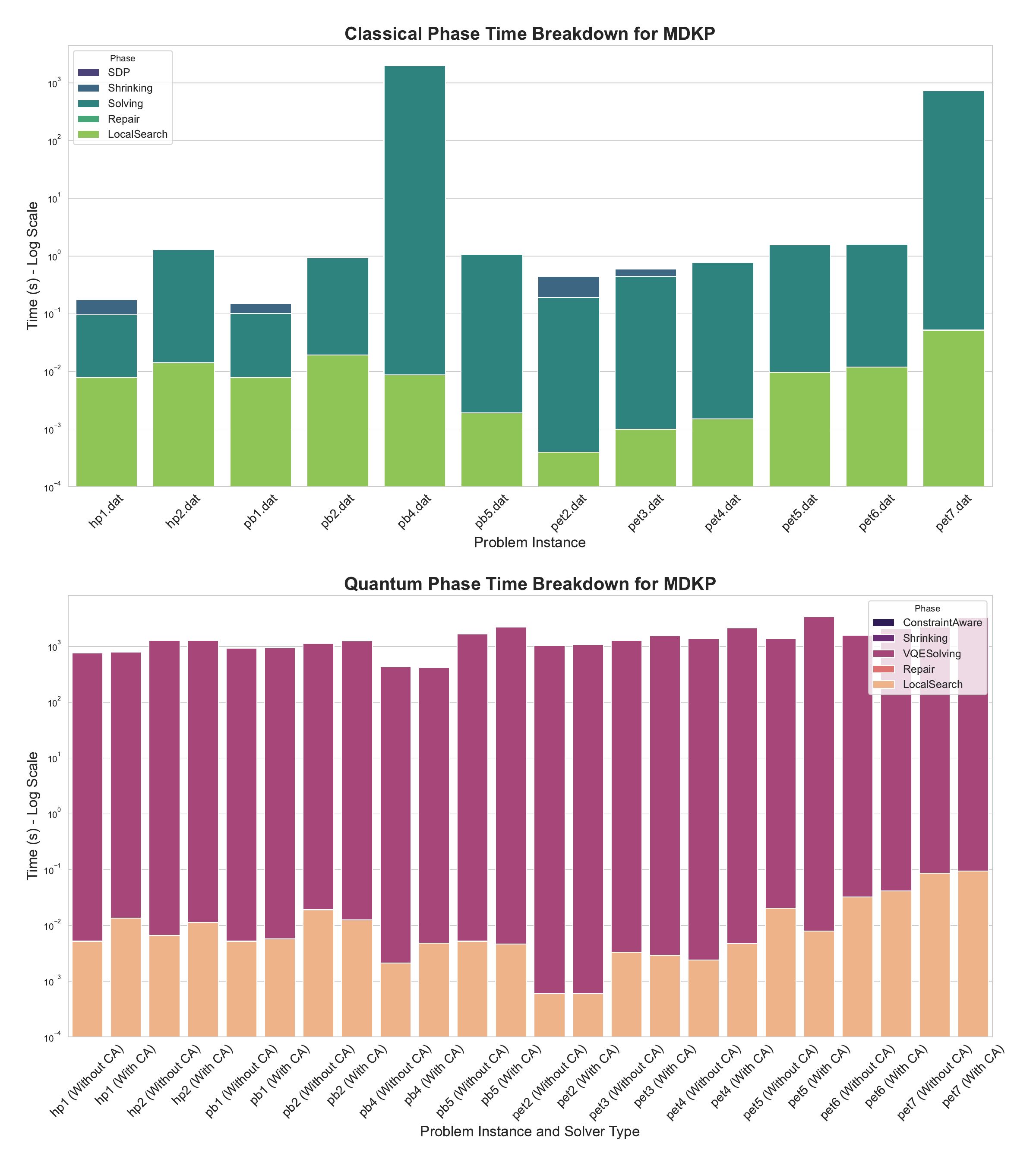}
    \caption{\label{fig:mdkp_phase_breakdown}Phase-wise runtime breakdown (log scale) for MDKP instances. Shrinking and solving phases dominate the runtime, with added overhead for constraint-aware strategies in the quantum pipeline.}
\end{figure*}

\begin{figure*}[htbp]
    \centering
    \includegraphics[width=0.9\textwidth]{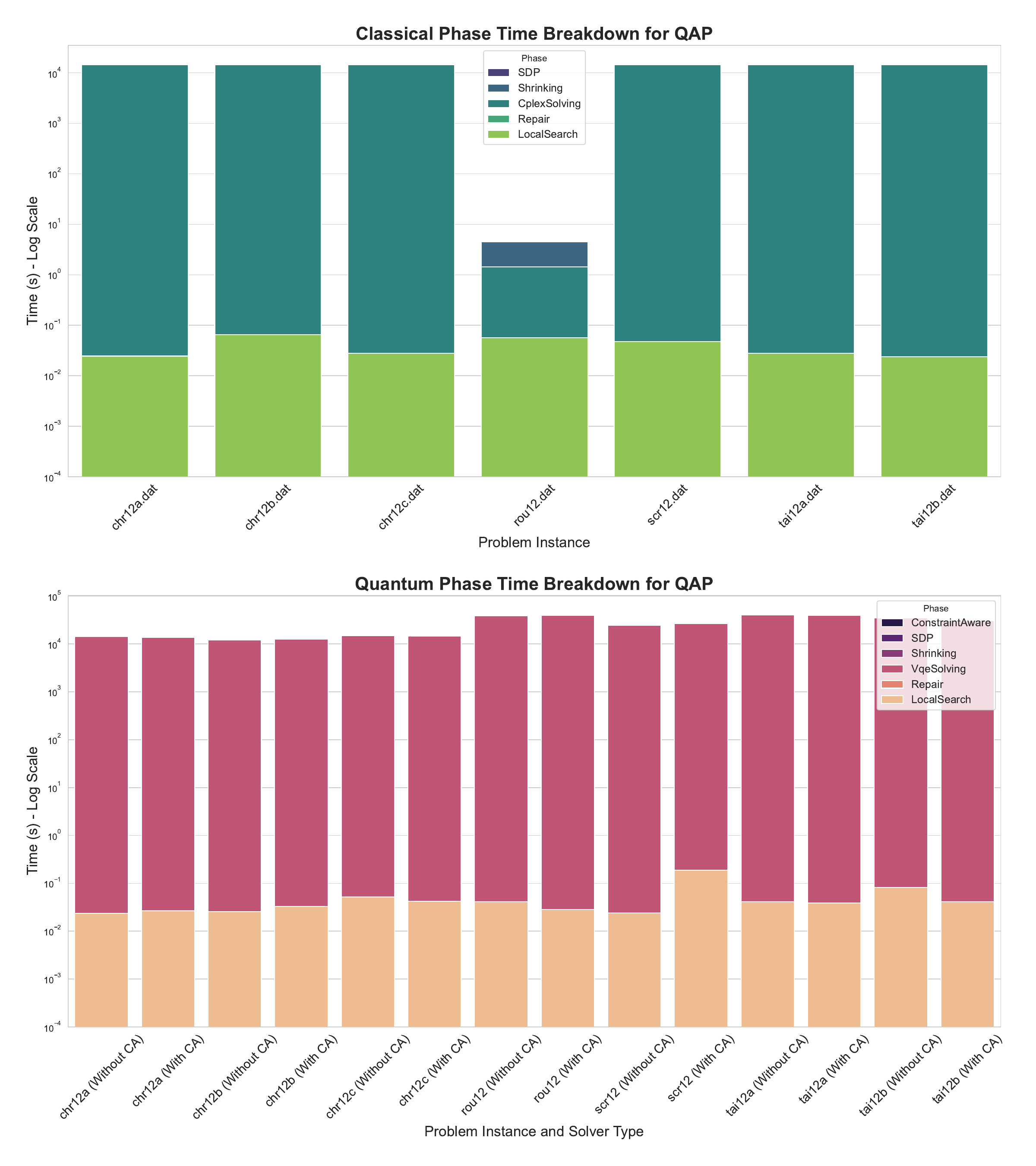}
    \caption{\label{fig:qap_phase_breakdown}Phase-wise runtime breakdown (log scale) for QAP instances. Due to the complexity of QAP, time spent solving is notably higher in the quantum pipeline. For the classical CPLEX, a time limit of 4 hours was set.}
\end{figure*}

\vspace{1em}
\noindent
To gain a deeper understanding of where computational resources are spent, we present a detailed \textbf{phase-wise runtime breakdown} for each problem category. Figure~\ref{fig:mis_phase_breakdown}, Figure~\ref{fig:mdkp_phase_breakdown}, and Figure~\ref{fig:qap_phase_breakdown} illustrate the breakdown (on a log scale) of total runtime per instance for the MIS, MDKP, and QAP benchmarks, respectively.

\medskip
\noindent
Each stacked bar corresponds to a benchmark instance, and decomposes the runtime into key phases for both classical and quantum pipelines:

\begin{itemize}
    \item \textbf{Classical Pipeline:}
    \begin{itemize}
        \item \textbf{SDP Correlation Computation} – Time taken to compute pairwise correlations via semidefinite programming.
        \item \textbf{Graph Shrinking} – Time for adaptive instance reduction based on the computed correlations.
        \item \textbf{CPLEX Solving} – Time to solve the shrunken problem using classical optimization (dominant component).
        \item \textbf{Repair and Local Search} – Postprocessing phases to improve feasibility or solution quality.
    \end{itemize}

    \item \textbf{Quantum Pipeline:}
    \begin{itemize}
        \item \textbf{Constraint-Aware Correlation Calculation} – Time to compute penalized correlations using constraint-weighted SDP or other tailored heuristics.
        \item \textbf{Graph Shrinking} – Same reduction framework applied to construct the QUBO-compatible problem.
        \item \textbf{VQE Solving} – Time to solve the QUBO using the Variational Quantum Eigensolver (dominant component).
        \item \textbf{Repair and Local Search} – Post-VQE phases to enhance feasibility or objective value.
    \end{itemize}
\end{itemize}

\medskip
\noindent
While the breakdown reveals some variability across problems and instances, several consistent patterns emerge:
\begin{itemize}
    \item In both classical and quantum pipelines, the \textbf{solving phase (CPLEX or VQE)} accounts for the vast majority of runtime — typically over \textbf{95\%} of the total wall-clock time.
    \item The \textbf{local search phase} is the second most time-consuming component, contributing around \textbf{2\%} of total runtime on average.
    \item For the quantum pipeline, the \textbf{constraint-aware correlation calculation} incurs a modest additional cost relative to the classical SDP correlations, due to the need to incorporate penalty-weighted structure into the shrinking process.
    \item Other phases — such as shrinking and QUBO construction — take negligible time in comparison, especially when viewed on a log scale.
\end{itemize}

\noindent
These results underscore that while the majority of runtime is dominated by solving and postprocessing, the design of the correlation computation and shrinking phases remains crucial for ensuring quality and feasibility — particularly in the quantum pipeline where additional structure is encoded upfront.

\section{Future Outlook \& Conclusion}
\label{sec:discussion}

In this work, we introduced a hybrid classical–quantum framework that leverages adaptive graph shrinking to enable scalable quantum optimization of classically challenging constrained combinatorial problems. Our method introduces a constraint-aware shrinking process, which carefully avoids merges that may compromise feasibility, and integrates a verification-and-repair pipeline to ensure valid solutions. By applying this framework to benchmark problems including MDKP, MIS, and QAP, we demonstrate notable improvements in solution feasibility and quality, especially when quantum hardware constraints impose tight limits on problem size.

Our extensive empirical analysis shows that although constraint-aware shrinking introduces additional computational overhead—primarily due to penalized correlation calculations—it significantly improves the quality and feasibility of quantum solutions. The solving phase (via VQE or CPLEX) remains the dominant contributor to runtime, yet the structural advantages conferred by constraint-aware preprocessing yield more recoverable and repair-efficient solutions. Furthermore, adaptive strategies for recalculating correlations and controlling the reduction process contribute to maintaining structural integrity throughout the shrinking phase.

\subsection*{Future Outlook}

Looking ahead, several directions emerge for advancing this framework:

\begin{itemize}
    \item \textbf{Integration with Noisy Quantum Hardware.} Future work should explore deploying this framework on actual quantum processors, with error mitigation strategies tailored to the reduced, constraint-aware subproblems.

    \item \textbf{Learning-Based Shrinking Heuristics.} Incorporating machine learning techniques, such as graph neural networks (GNNs), to predict merge candidates or adaptively learn penalization strategies could further enhance shrinking decisions beyond SDP-based heuristics.

    \item \textbf{Dynamic Penalty Adaptation.} The constraint-aware merge penalty parameter $\lambda$ was statically tuned in our experiments. Reinforcement learning or metaheuristic adaptation of this parameter could offer better generalization across problem classes.

    \item \textbf{Support for Larger Problem Classes.} Extending the framework to additional problem types—such as the Vehicle Routing Problem (VRP), Graph Coloring, or Supply Chain Design—can demonstrate its broader applicability and scalability.

    \item \textbf{Advanced Reconstruction Pipelines.} While our current repair methods are heuristic, integrating formal decoding mechanisms or probabilistic reconstruction could reduce information loss during the shrinking–solving–reconstruction cycle.

\end{itemize}

Overall, our work illustrates how hybrid architectures that combine classical preprocessing with quantum optimization can overcome current hardware limitations, and lays the foundation for principled, scalable quantum solvers for constrained combinatorial optimization problems.

\section{Acknowledgement}

This research is supported by the National Research Foundation, Singapore under its Quantum Engineering Programme 2.0 (NRF2021-QEP2-02-P01).

\clearpage
\bibliography{reference} 

\appendix

\onecolumngrid  

\section{Graph Shrinking Algorithm}

\begin{algorithm}[H]
\caption{Constraint-Aware Graph Shrinking with Reactive Solution Repair}
\label{alg:graph_shrinking_reactive_proactive}

\textbf{Input:} Weighted graph \( G = (V, E) \), SDP correlation matrix \( X \), target size \( k \), recalculation interval \( r \), penalty factor \( \lambda \), constraint-aware penalty function \( \Pi(\cdot, \cdot) \) \\
\textbf{Output:} Reduced graph \( G' \), partition \( \mathcal{P} \), final solution \( \mathcal{S} \)

\begin{algorithmic}[1]

\Procedure{Graph Shrinking (Proactive Phase)}{}
    \State Initialize \( G' \gets G \), node map \( M[v] \gets \text{index of } v \)
    \State Initialize partition map \( \mathcal{P}[v] \gets \text{None} \) for all \( v \in V \)
    \State Initialize step log \texttt{steps} as an empty list
    \State Initialize counters: \( t \gets 1 \), \( c \gets 0 \)
    
    \While{\( |V(G')| > k \)}
        \State Align \( X \) with current nodes: \( X \gets X[M, M] \)
        \State Determine candidate pairs and compute composite scores:
        \[
        S(C_i, C_j) = \mathbb{E}_{u \in C_i, v \in C_j} [X_{uv}] - \lambda \cdot \Pi(C_i, C_j)
        \]
        \State Select \( (i,j) = \arg\max_{(i,j)} S(C_i, C_j) \)
        \If{no valid merge candidate found}
            \State \textbf{break}
        \EndIf
        \State \( \sigma_{ij} \gets \text{sign}(X_{ij}) \)
        \State Append merge step \( (i, j, \sigma_{ij}, j) \) to \texttt{steps}

        \If{\( \sigma_{ij} > 0 \)}
            \State Assign \( \mathcal{P}[i], \mathcal{P}[j] \gets 0 \) or consistent with current assignment
        \Else
            \State Assign \( \mathcal{P}[i], \mathcal{P}[j] \) to opposite partitions
        \EndIf

        \State Merge node \( i \) into node \( j \): update edges of \( j \), remove \( i \) from \( G' \)
        \State Update index map \( M \), increment counters \( t \gets t+1, c \gets c+1 \)

        \If{\( c \geq r \)}
            \State Recompute SDP correlation matrix: \( X \gets \textsc{CalculateCorrelations}(\textsc{SolveSDP}(G')) \)
            \State Reset merge counter \( c \gets 0 \)
        \EndIf
    \EndWhile
    \State \textbf{return} \( G' \), \( \mathcal{P} \), \texttt{steps}
\EndProcedure

\Procedure{Solution Reconstruction + Repair (Reactive Phase)}{\texttt{steps}, $\mathcal{P}'$}
    \State Initialize reconstructed solution \( \mathcal{S} \gets \mathcal{P}' \)
    \For{each \( (i, j, \sigma_{ij}, j) \) in \texttt{steps} (reverse order)}
        \If{\( \sigma_{ij} > 0 \)}
            \State \( \mathcal{S}[i] \gets \mathcal{S}[j] \)
        \Else
            \State \( \mathcal{S}[i] \gets 1 - \mathcal{S}[j] \)
        \EndIf
    \EndFor
    \If{ \textsc{VerifySolution}(\( \mathcal{S} \)) = \texttt{False} }
        \State \( \mathcal{S} \gets \textsc{GreedyRepair}(\mathcal{S}) \)
    \EndIf
    \State \textbf{return} \( \mathcal{S} \)
\EndProcedure

\end{algorithmic}
\end{algorithm}

\section{Evaluation of Shrinking Ratios and Adaptive Reduction}
\label{app:shrinking_abalation}

We conducted an ablation study to investigate how the extent of graph shrinking influences performance across different problem instances. Specifically, we compared the following.

\begin{itemize}
    \item \textbf{Fixed Shrinking (2/3)}: Reducing the graph to two thirds of its original size.
    \item \textbf{Fixed Shrinking (1/2)}: Reducing the graph to half its original size.
    \item \textbf{Adaptive Shrinking}: Dynamically determining the amount of shrinking based on instance structure and intermediate spectral criteria.
\end{itemize}

The following tables report the effects of these strategies on execution time, solution quality (RSQ\% or optimality gap), and final feasibility. The results informed our decision to adopt \textbf{adaptive shrinking} for the main experiments.

\subsection{Results}

Tables~\ref{tab:results_mdkp} and~\ref{tab:results_mis} present the performance of our graph shrinking framework on benchmark instances of the Multidimensional Knapsack Problem (MDKP) and the Maximum Independent Set (MIS) problem, respectively. Our method consistently produced feasible solutions across all instances and achieved competitive optimality gaps. 

We compared three shrinking strategies: fixed ratio reductions to \(\lfloor 2/3 \cdot n \rfloor\) and \(\lfloor 1/2 \cdot n \rfloor\), and an adaptive strategy (Adap GS) that dynamically selects the final problem size based on spectral characteristics of the instance.

For baselines, we include results obtained from direct quantum solvers, such as variational quantum eigensolver (VQE), quantum approximation optimization algorithm (QAOA), and quantum random access optimization (QRAO), in the unshrunk QUBO formulations, and classical solvers such as CPLEX applied to the full LP formulations.

To assess quantum resource efficiency, Appendix~\ref{appendix:resource_usage} provides a detailed analysis of the number of qubits, ansatz depth, gate counts, trainable parameters, and total runtime required by each method. These metrics help characterize the scalability of quantum optimization methods under different preprocessing strategies.

\vspace{1em}

We compared our approach against the following baselines:

\begin{itemize}
    \item \textbf{Classical Solvers}: Each instance was solved using CPLEX, with known optimal solutions from benchmark datasets used as the ground truth.
    \item \textbf{Direct Quantum Solvers}: VQE, QAOA, and QRAO~\cite{fuller2024approximate} were applied directly to the original QUBO formulations without pre-processing. These results are included in Tables~\ref{tab:results_mdkp} and~\ref{tab:results_mis}.
\end{itemize}

\begin{table*}[h]
    \caption{\label{tab:results_mdkp}Performance of different graph shrinking ratios and direct quantum solvers on MDKP instances.}
    \begin{ruledtabular}
    \begin{tabular*}{\textwidth}{@{\extracolsep{\fill}}c|c|c|cc|cc|cc|c}
        \textbf{Instance} & \textbf{Optimal (Known)} & \textbf{Qubits (QUBO)} & 
        \multicolumn{2}{c|}{\textbf{2/3 GS}} & 
        \multicolumn{2}{c|}{\textbf{1/2 GS}} & 
        \multicolumn{2}{c|}{\textbf{Adap GS}} & 
        \textbf{VQE Gap (\%)} \\
        \cline{4-9}
        & & & \textbf{Size} & \textbf{Gap (\%)} & 
        \textbf{Size} & \textbf{Gap (\%)} & 
        \textbf{Size} & \textbf{Gap (\%)} & \\
        \hline
        hp1 & 3418 & 60 & 40 & 16.15 & 30 & 19.77 & 50 & 15.77 & 39.76 \\
        hp2 & 3186 & 67 & 44 & 34.65 & 34 & 34.74 & 57 & 16.61 & 12.34 \\
        pb1 & 3090 & 59 & 39 & 22.01 & 30 & 38.73 & 50 & 12.92 & 19.94 \\
        pb2 & 3186 & 66 & 44 & 26.37 & 34 & 24.45 & 56 & 11.49 & 19.49 \\
        pb4 & 95168 & 45 & 30 & 35.34 & 23 & 39.07 & 37 & 11.41 & inf. \\
        pb5 & 2139 & 116 & 77 & 19.07 & 59 & 25.19 & 95 & 12.53 & 4.25 \\
        pet2 & 87061 & 99 & 66 & 3.97 & 50 & 7.53 & 80 & 0.21 & 41.07 \\
        pet3 & 4015 & 102 & 68 & 19.43 & 52 & 28.76 & 82 & 2.49 & 4.98 \\
        pet4 & 6120 & 107 & 71 & 19.77 & 54 & 19.93 & 85 & 16.99 & 66.58 \\
        pet5 & 12400 & 122 & 81 & 23.23 & 62 & 18.50 & 98 & 12.66 & 33.23 \\
        pet6 & 10618 & 86 & 57 & 39.41 & 44 & 10.44 & 72 & 7.23 & 12.50 \\
        pet7 & 16537 & 100 & 66 & 30.54 & 50 & 35.49 & 85 & 10.52 & 43.46 \\
    \end{tabular*}
    \end{ruledtabular}
\end{table*}

\begin{table*}[h]
    \caption{\label{tab:results_mis}Performance of shrinking strategies and direct quantum solvers on MIS benchmark instances.}

    \begin{ruledtabular}
    \begin{tabular*}{\textwidth}{@{\extracolsep{\fill}}c|c|c|cc|cc|cc|c|c|c}
        \textbf{Instance} & \textbf{Optimal} & \textbf{Qubits} & 
        \multicolumn{2}{c|}{\textbf{2/3 GS}} & 
        \multicolumn{2}{c|}{\textbf{1/2 GS}} & 
        \multicolumn{2}{c|}{\textbf{Adap GS}} & 
        \textbf{VQE  (\%)} & \textbf{QAOA (\%)} & \textbf{QRAO  (\%)} \\
        \cline{4-9}
        & & & \textbf{Size} & \textbf{RSQ (\%)} & 
        \textbf{Size} & \textbf{RSQ (\%)} & 
        \textbf{Size} & \textbf{RSQ (\%)} & & & \\
        \hline
        1dc.64 & 10 & 64 & 34 & 80.00 & 32 & 60.0 & 51 & 90.0 & 87.5 & M.E & inf \\
        1tc.16 & 8 & 16 & 12 & 87.5 & 8 & 75.0 & 9 & 87.5 & 75.0 & 100.0 & 87.5 \\
        1tc.32 & 12 & 32 & 22 & 83.33 & 17 & 66.67 & 20 & 100.0 & 75.0 & 83.3 & 58.3 \\
        1et.64 & 18 & 64 & 42 & 77.78 & 32 & 77.78 & 46 & 88.89 & 77.8 & inf & inf \\
        1tc.64 & 20 & 64 & 44 & 77.78 & 32 & 75.00  & 43 & 90.0 & 40.0 & 50.0 & 40.0 \\
        1tc.8 & 4 & 8 & 6 & 100.0 & 4 & 100.0 & 4 & 100.0 & 100.0 & 100.0 & 100.0 \\
    \end{tabular*}
    \end{ruledtabular}
\end{table*}

\subsection{Observations}

Our graph shrinking framework significantly reduced problem sizes while maintaining high-quality solutions. Fixed-ratio strategies reduced the problem to \(\lfloor 2/3 \cdot n \rfloor\) and \(\lfloor 1/2 \cdot n \rfloor\) variables (qubits), where \(n\) is the original problem size. Additionally, the Adaptive Graph Shrinking (Adap GS) approach dynamically determined the final reduced size using spectral gap heuristics, allowing instance-specific adaptation.

The results in Tables~\ref{tab:results_mdkp} and~\ref{tab:results_mis} demonstrate that:
\begin{itemize}
    \item \textbf{All shrinking strategies consistently produced feasible solutions}, even when direct quantum solvers encountered infeasibility due to qubit limits or memory errors.
    \item \textbf{QAOA and QRAO were only applicable to MIS instances}, due to the memory-intensive nature of MDKP. VQE remained the only viable quantum baseline for MDKP.
\end{itemize}

We further summarize key observations:

\begin{itemize}
    \item \textbf{Fixed Shrinking (\(\lfloor 2/3 \cdot n \rfloor\) and \(\lfloor 1/2 \cdot n \rfloor\))}: While reducing the problem size substantially, the more aggressive 1/2 reduction often resulted in higher optimality gaps. The 2/3 reduction balanced runtime and solution quality more favorably.
    
    \item \textbf{Adaptive Shrinking (Adap GS)}: By leveraging spectral analysis, this method produced the best trade-off between size reduction and optimality. It consistently maintained low gaps while keeping the problem small enough for tractable quantum optimization.
\end{itemize}

\section{Complete Tabular Results}
\label{app:table}

This section presents complete experimental results on all benchmark instances from the \textbf{Multi-Dimensional Knapsack Problem (MDKP)}, \textbf{Maximum Independent Set (MIS)}, and \textbf{Quadratic Assignment Problem (QAP)}, using \textbf{adaptive graph shrinking}.

For each instance, we first applied the adaptive shrinking procedure to reduce the problem size, and then solved the resulting smaller problem using two approaches:
\begin{itemize}
    \item \textbf{Classical Solver (CPLEX)}: Applied to the adaptively shrunken instance. All CPLEX results use the non constraint-aware formulation.
    \item \textbf{Variational Quantum Eigensolver (VQE)}: Applied to both constraint-aware and non-constraint-aware versions of the same shrunken instance.
\end{itemize}

The tables report solution quality (optimality gap or relative solution quality), feasibility, and runtime. These results provide insight into how classical and quantum solvers perform under graph shrinking and constraint modeling choices.

\begin{table*}[h]
    \caption{\label{tab:mis-cplex}Performance of MIS instances using CPLEX after adaptive graph shrinking.}
    \begin{ruledtabular}
    \begin{tabular*}{\textwidth}{@{\extracolsep{\fill}}lrrrrrlr}
        \textbf{Instance} & \textbf{Initial Size} & \textbf{Final Size} & \textbf{Final Objective} & \textbf{Feasible} & \textbf{Total Time (s)} & \textbf{RSQ (\%)} \\
        \hline
        1tc.16.txt & 16 & 9  & 8  & True & 0.0736 & 100 \\
        1dc.64.txt & 64 & 51 & 8  & True & 3.2339 & 80  \\
        1et.64.txt & 64 & 46 & 18 & True & 0.5083 & 100 \\
        1tc.8.txt  & 8  & 4  & 4  & True & 0.0283 & 100 \\
        1tc.32.txt & 32 & 20 & 12 & True & 0.1215 & 100 \\
        1tc.64.txt & 64 & 43 & 20 & True & 0.4415 & 100 \\
    \end{tabular*}
    \end{ruledtabular}
\end{table*}

\begin{table*}[h]

    \caption{Performance of MDKP instances using CPLEX}
    \label{tab:mdkp-cplex}
    \begin{ruledtabular}
    \begin{tabular*}{\textwidth}{@{\extracolsep{\fill}}lrrrrrlr}
    \textbf{Instance} & \textbf{Initial Size} & \textbf{Final Size} & \textbf{Final Objective} & \textbf{Feasible} & \textbf{Total Time (s)} & \textbf{Opt Gap (\%)} \\
    \hline
hp1.dat    & 60  & 50 & 3329  & True & 0.4277   & 2.60  \\
hp2.dat    & 67  & 57 & 2828  & True & 2.0039   & 11.24 \\
pb1.dat    & 59  & 50 & 2955  & True & 0.4066   & 4.37  \\
pb2.dat    & 66  & 56 & 2856  & True & 1.5821   & 10.36 \\
pb4.dat    & 45  & 37 & 86112 & True & 2004.19  & 9.51  \\
pb5.dat    & 116 & 95 & 1871  & True & 2.2447   & 12.53 \\
pet2.dat    & 99 & 80 & 86875  & True & 0.8644   & 0.21 \\
pet3.dat    & 102 & 82 & 4015  & True & 1.3155   & 0.00 \\
pet4.dat    & 107 & 85 & 6090  & True & 1.7866   & 0.49 \\
pet5.dat    & 122 & 98 & 12400  & True & 3.0364   & 0.08 \\
pet6.dat    & 86 & 72 & 10618  & True & 3.0565   & 14.58 \\
pet7.dat    & 100 & 85 & 16537  & True & 737.5989   & 14.22 \\

    \end{tabular*}
    \end{ruledtabular}
\end{table*}

\begin{table*}[h]
    \caption{\label{tab:mis-vqe}Performance of MIS instances using VQE (non-constraint-aware) after adaptive graph shrinking.}
    \begin{ruledtabular}
    \begin{tabular*}{\textwidth}{@{\extracolsep{\fill}}lcrrrrlr}
        \textbf{Instance} & \textbf{Constraint-Aware} & \textbf{Initial Size} & \textbf{Final Size} & \textbf{Final Objective} & \textbf{Feasible} & \textbf{Total Time (s)} & \textbf{RSQ (\%)} \\
        \hline
        1tc.16.txt & False & 16 & 9  & 7  & True & 48.6189 & 87.5 \\
        1dc.64.txt & False & 64 & 51 & 7  & True & 535.9980 & 70.0 \\
        1et.64.txt & False & 64 & 46 & 16 & True & 1874.3088 & 88.89 \\
        1tc.8.txt  & False & 8  & 4  & 4  & True & 9.6379 & 100.0 \\
        1tc.32.txt & False & 32 & 20 & 11 & True & 381.3125 & 91.67 \\
        1tc.64.txt & False & 64 & 43 & 18 & True & 1307.3168 & 90.0 \\
    \end{tabular*}
    \end{ruledtabular}
\end{table*}

\begin{table*}[h]
    \caption{\label{tab:mdkp-vqe}Performance of MDKP instances using VQE (non-constraint-aware) after adaptive graph shrinking.}
    \begin{ruledtabular}
    \begin{tabular*}{\textwidth}{@{\extracolsep{\fill}}lcrrrrlr}
        \textbf{Instance} & \textbf{Constraint-Aware} & \textbf{Initial Size} & \textbf{Final Size} & \textbf{Final Objective} & \textbf{Feasible} & \textbf{Total Time (s)} & \textbf{Optimality Gap (\%)} \\
        \hline
        hp1.dat    & False & 60  & 50 & 2881  & True & 768.8991   & 15.77 \\
        hp2.dat    & False & 67  & 57 & 2655  & True & 1288.7103  & 16.61 \\
        pb1.dat    & False & 59  & 50 & 2530  & True & 934.3716   & 18.12 \\
        pb2.dat    & False & 66  & 56 & 2820  & True & 1139.3174  & 11.49 \\
        pb4.dat    & False & 45  & 37 & 80965 & True & 431.4437   & 14.92 \\
        pb5.dat    & False & 116 & 95 & 1871  & True & 1674.3447  & 12.53 \\
        pet2.dat   & False & 99  & 80 & 86875 & True & 1031.4054  & 0.21 \\
        pet3.dat   & False & 102 & 82 & 3330  & True & 1283.2781  & 17.06 \\
        pet4.dat   & False & 107 & 85 & 6090  & True & 1385.6190  & 16.99 \\
        pet5.dat   & False & 122 & 98 & 12400 & True & 1388.4558  & 12.66 \\
        pet6.dat   & False & 86  & 72 & 10618 & True & 1581.9467  & 7.23 \\
        pet7.dat   & False & 100 & 85 & 16537 & True & 2221.0501  & 10.52 \\
    \end{tabular*}
    \end{ruledtabular}
\end{table*}

\begin{table*}[h]
    \caption{\label{tab:mis-vqe-constraint}Performance of MIS instances using VQE (constraint-aware) after adaptive graph shrinking.}
    \begin{ruledtabular}
    \begin{tabular*}{\textwidth}{@{\extracolsep{\fill}}lcrrrrlr}
        \textbf{Instance} & \textbf{Constraint-Aware} & \textbf{Initial Size} & \textbf{Final Size} & \textbf{Final Objective} & \textbf{Feasible} & \textbf{Total Time (s)} & \textbf{RSQ (\%)} \\
        \hline
        1tc.16.txt & True & 16 & 9  & 7  & True & 56.1805  & 87.5 \\
        1dc.64.txt & True & 64 & 51 & 7  & True & 499.8730 & 90.0 \\
        1et.64.txt & True & 64 & 46 & 16 & True & 1249.4107 & 88.89 \\
        1tc.8.txt  & True & 8  & 4  & 4  & True & 4.1182   & 100.0 \\
        1tc.32.txt & True & 32 & 20 & 12 & True & 259.8223 & 100.0 \\
        1tc.64.txt & True & 64 & 43 & 19 & True & 1097.3758 & 95.0 \\
    \end{tabular*}
    \end{ruledtabular}
\end{table*}

\begin{table*}[h]
    \caption{\label{tab:mdkp-vqe-constraint}Performance of MDKP instances using VQE (constraint-aware) after adaptive graph shrinking.}
    \begin{ruledtabular}
    \begin{tabular*}{\textwidth}{@{\extracolsep{\fill}}lcrrrrlr}
        \textbf{Instance} & \textbf{Constraint-Aware} & \textbf{Initial Size} & \textbf{Final Size} & \textbf{Final Objective} & \textbf{Feasible} & \textbf{Total Time (s)} & \textbf{Optimality Gap (\%)} \\
        \hline
        hp1.dat    & True & 60  & 50 & 3221   & True & 793.2083   & 5.76  \\
        hp2.dat    & True & 67  & 57 & 2873   & True & 1296.4207  & 9.82  \\
        pb1.dat    & True & 59  & 50 & 2691   & True & 955.1857   & 12.91 \\
        pb2.dat    & True & 66  & 56 & 2940   & True & 1265.4001  & 7.81  \\
        pb4.dat    & True & 45  & 37 & 84306  & True & 420.8970   & 11.41 \\
        pb5.dat    & True & 116 & 95 & 1871   & True & 2219.8709  & 12.53 \\
        pet2.dat   & True & 99  & 80 & 86875  & True & 1031.4054  & 0.21  \\
        pet3.dat   & True & 102 & 82 & 3915   & True & 1283.2781  & 2.49 \\
        pet4.dat   & True & 107 & 85 & 6090   & True & 1385.6190  & 2.29  \\
        pet5.dat   & True & 122 & 98 & 12400  & True & 1388.4558  & 14.11 \\
        pet6.dat   & True & 86  & 72 & 10618  & True & 1581.9467  & 2.42  \\
        pet7.dat   & True & 100 & 85 & 16537  & True & 2221.0501  & 2.49  \\
    \end{tabular*}
    \end{ruledtabular}
\end{table*}

\begin{table*}[h]
    \caption{\label{tab:qap-cplex}Performance of QAP instances using CPLEX after adaptive graph shrinking.}
    \begin{ruledtabular}
    \begin{tabular*}{\textwidth}{@{\extracolsep{\fill}}lrrrlrrr}
        \textbf{Instance} & \textbf{Initial Size} & \textbf{Final Size} & \textbf{Final Objective} & \textbf{Feasible} & \textbf{Total Time (s)} & \textbf{Known Optimal} & \textbf{Optimality Gap (\%)} \\
        \hline
        chr12a.dat & 144 & 121 & 17272     & True & 14420.71 & 9552     & 80.82 \\
        chr12b.dat & 144 & 120 & 15202     & True & 14425.77 & 9742     & 56.05 \\
        chr12c.dat & 144 & 123 & 16292     & True & 14416.32 & 11156    & 46.04 \\
        rou12.dat  & 144 & 128 & 254074    & True & 15.69    & 235528   & 7.87  \\
        scr12.dat  & 144 & 121 & 32976     & True & 14415.24 & 31410    & 4.99  \\
        tai12a.dat & 144 & 128 & 251534    & True & 14442.63 & 224416   & 12.08 \\
        tai12b.dat & 144 & 126 & 49653413  & True & 14443.91 & 39464925 & 25.82 \\
    \end{tabular*}
    \end{ruledtabular}
\end{table*}

\begin{table*}[h]
    \caption{\label{tab:qap-vqe}Performance of QAP instances using VQE with and without constraint-aware shrinking.}
    \begin{ruledtabular}
    \begin{tabular*}{\textwidth}{@{\extracolsep{\fill}}lcccccrrr}
        \textbf{Instance} & \textbf{Constraint-Aware} & \textbf{Initial Size} & \textbf{Final Size} & \textbf{Final Objective} & \textbf{Feasible} & \textbf{Time (s)} & \textbf{Known Optimal} & \textbf{Gap (\%)} \\
        \hline
        chr12a.dat & False & 144 & 121 & 15320    & True & 14320.06 & 9552     & 60.39 \\
        chr12b.dat & False & 144 & 120 & 15202    & True & 12062.80 & 9742     & 56.05 \\
        chr12c.dat & False & 144 & 123 & 17830    & True & 14792.90 & 11156    & 59.82 \\
        rou12.dat  & False & 144 & 128 & 254224   & True & 38976.71 & 235528   & 7.94  \\
        scr12.dat  & False & 144 & 121 & 318840   & True & 24494.45 & 31410    & 1.51  \\
        tai12a.dat & False & 144 & 128 & 246826   & True & 40483.81 & 224416   & 9.99  \\
        tai12b.dat & False & 144 & 126 & 45510624 & True & 35329.84 & 39464925 & 15.32 \\
        chr12a.dat & True  & 144 & 121 & 13372    & True & 13624.64 & 9552     & 39.99 \\
        chr12b.dat & True  & 144 & 120 & 10102    & True & 12630.87 & 9742     & 3.70  \\
        chr12c.dat & True  & 144 & 123 & 14866    & True & 14674.41 & 11156    & 33.26 \\
        rou12.dat  & True  & 144 & 128 & 240038   & True & 39806.68 & 235528   & 3.70  \\
        scr12.dat  & True  & 144 & 121 & 32976    & True & 26717.80 & 31410    & 4.99  \\
        tai12a.dat & True  & 144 & 128 & 252116   & True & 30903.54 & 224416   & 10.27 \\
    \end{tabular*}
    \end{ruledtabular}
\end{table*}

\clearpage
\section{\label{appendix:resource_usage}Quantum Algorithm Metrics}

\begin{table*}[h]
    \caption{Resource Usage by Instance and Approach. Each instance's resources are detailed for VQE, QAOA, QRAO, 1/2 GS, Adaptive GS, and 2/3 GS approaches.}
    \centering
    \footnotesize
    \begin{ruledtabular}
    \begin{tabular}{lcccccc}
        \textbf{Instance} & \textbf{Approach} & \textbf{Qubits} & \textbf{Depth} & \textbf{Gate Count} & \textbf{2-Qubit Gates} & \textbf{Parameters} \\
        \hline
        \textbf{1dc.64} 
            & VQE         & 50 & 61  & 547  & 147  & 400 \\
            & QAOA        & 50 & 252 & 1377 & 818  & 559 \\
            & QRAO        & 18 & 25  & 142  & 34   & 108 \\
            & 1/2 GS      & 26 & 45  & 437  & 125  & 312 \\
            & Adaptive GS & 39 & 58  & 658  & 190  & 468 \\ 
            & 2/3 GS      & 34 & 53  & 573  & 165  & 408 \\ 
        \hline
        \textbf{1tc.16} 
            & VQE         & 16 & 27  & 173  & 45   & 128 \\
            & QAOA        & 16 & 30  & 114  & 44   & 70  \\
            & QRAO        & 6  & 13  & 54   & 10   & 34  \\
            & 1/2 GS      & 9  & 28  & 148  & 40   & 108 \\
            & Adaptive GS & 10 & 29  & 165  & 45   & 120 \\ 
            & 2/3 GS      & 12 & 31  & 199  & 55   & 144 \\ 
        \hline
        \textbf{1tc.32} 
            & VQE         & 32 & 43  & 349  & 93   & 256 \\
            & QAOA        & 32 & 54  & 300  & 136  & 164 \\
            & QRAO        & 13 & 20  & 102  & 24   & 78  \\
            & 1/2 GS      & 17 & 36  & 284  & 80   & 204 \\
            & Adaptive GS & 21 & 40  & 352  & 100  & 252 \\ 
            & 2/3 GS      & 22 & 41  & 369  & 105  & 264 \\ 
        \hline
        \textbf{1et.64} 
            & VQE         & 62 & 73  & 679  & 183  & 496 \\
            & QAOA        & 62 & 105 & 978  & 528  & 450 \\
            & QRAO        & 24 & 31  & 190  & 46   & 144 \\
            & 1/2 GS      & 32 & 51  & 539  & 155  & 384 \\
            & Adaptive GS & 45 & 64  & 760  & 220  & 540 \\ 
            & 2/3 GS      & 42 & 61  & 709  & 205  & 504 \\ 
        \hline
        \textbf{1tc.64} 
            & VQE         & 64 & 75  & 701  & 189  & 512 \\
            & QAOA        & 64 & 105 & 768  & 384  & 384 \\
            & QRAO        & 23 & 30  & 182  & 44   & 138 \\
            & 1/2 GS      & 33 & 52  & 556  & 160  & 396 \\
            & Adaptive GS & 43 & 62  & 726  & 210  & 516 \\ 
            & 2/3 GS      & 44 & 63  & 743  & 215  & 528 \\ 
        \hline
        \textbf{1tc.8} 
            & VQE         & 8  & 19  & 85   & 21   & 64  \\
            & QAOA        & 8  & 12  & 42   & 12   & 30  \\
            & QRAO        & 4  & 11  & 30   & 6    & 24  \\
            & 1/2 GS      & 5  & 24  & 80   & 20   & 60  \\
            & Adaptive GS & 5  & 24  & 80   & 20   & 60  \\ 
            & 2/3 GS      & 6  & 25  & 97   & 25   & 72  \\ 
    \end{tabular}
    \end{ruledtabular}
    \label{tab:resource_usage_1}
\end{table*}

\begin{table*}[p]
    \caption{Resource Usage by Instance and Approach. Each instance's resources are detailed for VQE, 1/2 GS, Adaptive GS, and 2/3 GS approaches.}
    \centering
    \footnotesize
    \begin{ruledtabular}
    \begin{tabular}{lcccccc}
        \textbf{Instance} & \textbf{Approach} & \textbf{Qubits} & \textbf{Depth} & \textbf{Gate Count} & \textbf{2-Qubit Gates} & \textbf{Parameters} \\
        \hline
        \textbf{hp1} 
            & VQE          & 60 & 75 & 836 & 236 & 600 \\
            & 1/2 GS       & 31 & 50 & 522 & 150 & 372 \\
            & Adaptive GS  & 37 & 56 & 624 & 180 & 444 \\
            & 2/3 GS       & 41 & 60 & 692 & 200 & 492 \\
        \hline
        \textbf{hp2} 
            & VQE          & 67 & 82 & 934 & 264 & 670 \\
            & 1/2 GS       & 34 & 53 & 573 & 165 & 408 \\
            & Adaptive GS  & 41 & 60 & 692 & 200 & 492 \\
            & 2/3 GS       & 46 & 65 & 777 & 225 & 552 \\
        \hline
        \textbf{pb1} 
            & VQE          & 59 & 74 & 822 & 232 & 590 \\
            & 1/2 GS       & 30 & 49 & 505 & 145 & 360 \\
            & Adaptive GS  & 36 & 55 & 607 & 175 & 432 \\
            & 2/3 GS       & 40 & 59 & 675 & 195 & 480 \\
        \hline
        \textbf{pb2} 
            & VQE          & 66 & 81 & 920 & 260 & 660 \\
            & 1/2 GS       & 34 & 53 & 573 & 165 & 408 \\
            & Adaptive GS  & 42 & 61 & 709 & 205 & 504 \\
            & 2/3 GS       & 45 & 60 & 626 & 176 & 450 \\
        \hline
        \textbf{pb4} 
            & VQE          & 45 & 60 & 626 & 176 & 450 \\
            & 1/2 GS       & 23 & 42 & 386 & 110 & 276 \\
            & Adaptive GS  & 31 & 50 & 522 & 150 & 372 \\
            & 2/3 GS       & 31 & 46 & 430 & 120 & 310 \\
        \hline
        \textbf{pb5} 
            & VQE          & 116 & 131 & 1620 & 460 & 1160 \\
            & 1/2 GS       & 59  & 78  & 998  & 290 & 708  \\
            & Adaptive GS  & 50  & 69  & 845  & 245 & 600  \\
            & 2/3 GS       & 78  & 93  & 1088 & 308 & 780  \\
        \hline
        \textbf{pet2} 
            & VQE          & 99 & 114 & 1382 & 392 & 990 \\
            & 1/2 GS       & 50 & 69  & 845  & 245 & 600 \\
            & Adaptive GS  & 40 & 59  & 675  & 195 & 480 \\
            & 2/3 GS       & 67 & 82  & 934  & 264 & 670 \\
        \hline
        \textbf{pet3} 
            & VQE          & 102 & 117 & 1424 & 404 & 1020 \\
            & 1/2 GS       & 52  & 71  & 879  & 255 & 624  \\
            & Adaptive GS  & 40  & 59  & 675  & 195 & 480 \\
            & 2/3 GS       & 69  & 84  & 962  & 272 & 690 \\
        \hline
        \textbf{pet4} 
            & VQE          & 107 & 122 & 1494 & 424 & 1070 \\
            & 1/2 GS       & 54  & 73  & 913  & 265 & 648 \\
            & Adaptive GS  & 40  & 59  & 675  & 195 & 480 \\
            & 2/3 GS       & 72  & 87  & 1004 & 284 & 720 \\
        \hline
        \textbf{pet5} 
            & VQE          & 122 & 137 & 1704 & 484 & 1220 \\
            & 1/2 GS       & 62  & 81  & 1049 & 305 & 704 \\
            & Adaptive GS  & 50  & 69  & 845  & 245 & 600 \\
            & 2/3 GS       & 82  & 97  & 1144 & 324 & 820 \\
        \hline
        \textbf{pet6} 
            & VQE          & 86 & 101 & 1210 & 340 & 860 \\
            & 1/2 GS       & 44 & 63  & 743  & 215 & 528 \\
            & Adaptive GS  & 46 & 65  & 777  & 225 & 552 \\
            & 2/3 GS       & 58 & 73  & 808  & 228 & 580 \\
        \hline
        \textbf{pet7} 
            & VQE          & 100 & 115 & 1396 & 396 & 1000 \\
            & 1/2 GS       & 51  & 70  & 862  & 250 & 612 \\
            & Adaptive GS  & 54  & 73  & 913  & 265 & 648 \\
            & 2/3 GS       & 68  & 83  & 948  & 268 & 680 \\
    \end{tabular}
    \end{ruledtabular}
    \label{tab:resource_usage_2}
\end{table*}

\end{document}